\documentclass[12pt]{article}
\usepackage{graphicx}
\usepackage{lscape}
\usepackage{citesort}

\begin{document}

\def\ds{\displaystyle}
\def\beq{\begin{equation}}
\def\eeq{\end{equation}}
\def\bea{\begin{eqnarray}}
\def\eea{\end{eqnarray}}
\def\beeq{\begin{eqnarray}}
\def\eeeq{\end{eqnarray}}
\def\ve{\vert}
\def\vel{\left|}
\def\ver{\right|}
\def\nnb{\nonumber}
\def\ga{\left(}
\def\dr{\right)}
\def\aga{\left\{}
\def\adr{\right\}}
\def\lla{\left<}
\def\rra{\right>}
\def\rar{\rightarrow}
\def\nnb{\nonumber}
\def\la{\langle}
\def\ra{\rangle}
\def\ba{\begin{array}}
\def\ea{\end{array}}
\def\tr{\mbox{Tr}}
\def\ssp{{\Sigma^{*+}}}
\def\sso{{\Sigma^{*0}}}
\def\ssm{{\Sigma^{*-}}}
\def\xis0{{\Xi^{*0}}}
\def\xism{{\Xi^{*-}}}
\def\qs{\la \bar s s \ra}
\def\qu{\la \bar u u \ra}
\def\qd{\la \bar d d \ra}
\def\qq{\la \bar q q \ra}
\def\gGgG{\la g^2 G^2 \ra}
\def\q{\gamma_5 \not\!q}
\def\x{\gamma_5 \not\!x}
\def\g5{\gamma_5}
\def\sb{S_Q^{cf}}
\def\sd{S_d^{be}}
\def\su{S_u^{ad}}
\def\ss{S_s^{??}}
\def\sbp{{S}_Q^{'cf}}
\def\sdp{{S}_d^{'be}}
\def\sup{{S}_u^{'ad}}
\def\ssp{{S}_s^{'??}}
\def\sig{\sigma_{\mu \nu} \gamma_5 p^\mu q^\nu}
\def\fo{f_0(\frac{s_0}{M^2})}
\def\ffi{f_1(\frac{s_0}{M^2})}
\def\fii{f_2(\frac{s_0}{M^2})}
\def\O{{\cal O}}
\def\sl{{\Sigma^0 \Lambda}}
\def\es{\!\!\! &=& \!\!\!}
\def\ap{\!\!\! &\approx& \!\!\!}
\def\ar{&+& \!\!\!}
\def\ek{&-& \!\!\!}
\def\kek{\!\!\!&-& \!\!\!}
\def\cp{&\times& \!\!\!}
\def\se{\!\!\! &\simeq& \!\!\!}
\def\eqv{&\equiv& \!\!\!}
\def\kpm{&\pm& \!\!\!}
\def\kmp{&\mp& \!\!\!}


\def\simlt{\stackrel{<}{{}_\sim}}
\def\simgt{\stackrel{>}{{}_\sim}}


\title{
         {\Large
                 {\bf
Polarization effects in exclusive semileptonic
$\Lambda_b \rar \Lambda \ell^+ \ell^-$ decay
                 }
         }
      }

\author{\vspace{1cm}\\
{\small T. M. Aliev \thanks
{e-mail: taliev@metu.edu.tr}~\footnote{permanent address:Institute
of Physics,Baku,Azerbaijan}\,\,,
M. Savc{\i} \thanks
{e-mail: savci@metu.edu.tr}} \\
{\small Physics Department, Middle East Technical University,
06531 Ankara, Turkey} }

\date{}

\begin{titlepage}
\maketitle
\thispagestyle{empty}

\begin{abstract}
The independent helicity amplitudes in the $\Lambda_b \rar \Lambda \ell^+
\ell^-$ decay in the standard model and its minimal extension, i.e., with
the new vector type interactions, are calculated. We calculate various
asymmetry parameters characterizing the angular dependence of the
differential decay width for the cascade decay $\Lambda_b \rar
\Lambda(\rar a+b) \, V^\ast (\rar \ell^+ \ell^-)$ with polarized and 
unpolarized heavy baryons. The sensitivity of the asymmetry parameters 
to the new Wilson coefficients are analyzed.
\end{abstract}

~~~PACS numbers: 12.60.--i, 13.30.--a. 13.88.+e
\end{titlepage}

\section{Introduction}

Rare B--decays induced by the flavor--changing neutral current (FCNC) 
$b \rar s$ or $b \rar d$ transitions occur at loop level in the standard 
model (SM), since FCNC transitions that are forbidden in the SM at tree 
level provide consistency check of the SM at quantum level.
These decays induced by the FCNC are also very promising tools for
establishing new physics beyond the SM. New physics appear in rare
decays through the Wilson coefficients which can take values different from
their SM counterpart or through the new operator structures in an effective
Hamiltonian (see for example \cite{R7101}--\cite{R7113}).

Among the hadronic, leptonic and semileptonic decays, the last decay channels 
are very significant, since they are theoretically, 
more or less, clean, and they have relatively larger branching ratio. 
The semileptonic decay channels is described by the 
$b \rar s(d) \ell^+ \ell^-$ transition and they
contain many observables like forward--backward asymmetry ${\cal A}_{FB}$,
lepton polarization asymmetries, etc. Existence of these observables is very
useful and serve as a testing ground for the SM and in
looking for new physics beyond th SM. For this reason, many processes, like
$B \rar \pi(\rho) \ell^+ \ell^-$ \cite{R7114},
$B \rar \ell^+ \ell^- \gamma$ \cite{R7115},
$B \rar K \ell^+ \ell^-$ \cite{R7116} and
$B \rar K^\ast \ell^+ \ell^-$
\cite{R7117,R7118,R7119,R7120,R7121,R7122,R7123,R7124} have been studied 
comprehensively. 

Recently, BELLE and BaBar Collaborations announced the
following results for the branching ratios of the
$B \rar K^\ast \ell^+ \ell^-$ and $B \rar K \ell^+ \ell^-$ decays:
\bea
{\cal B}(B \rar K^\ast \ell^+ \ell^-) = \left\{ \begin{array}{lc}
\left( 11.5^{+2.6}_{-2.4} \pm 0.8 \pm 0.2\right) \times
10^{-7}& \cite{R7125}~,\\ \\
\left( 0.78^{+0.19}_{-0.17} \pm 0.12\right) \times
10^{-6}& \cite{R7126}~,\end{array} \right. \nnb
\eea
\bea
{\cal B}(B \rar K \ell^+ \ell^-) = \left\{ \begin{array}{lc}
\left( 4.8^{+1.0}_{-0.9} \pm 0.3 \pm 0.1\right) \times
10^{-7}& \cite{R7125}~,\\ \\
\left( 0.34  \pm 0.07 \pm 0.12\right) \times
10^{-6}& \cite{R7126}~.\end{array} \right. \nnb
\eea
Another exclusive decay which is described at inclusive level by the 
$b \rar s \ell^+ \ell^-$ transition is the baryonic 
$\Lambda_b \rar \Lambda \ell^+ \ell^-$ decay. Unlike mesonic decays, the
baryonic decays 
could maintain the helicity structure of the effective Hamiltonian for 
the $b \rar s$ transition \cite{R7127}. Radiative and semileptonic decays of
$\Lambda_b$ such as $\Lambda_b \rar \Lambda \gamma$, $\Lambda_b \rar
\Lambda_c \ell \bar{\nu}_\ell$, $\Lambda_b \rar \Lambda \ell^+ \ell^-$
$(\ell = e, \mu, \tau)$ and $\Lambda_b \rar \Lambda \nu \bar{\nu}$ have been
extensively studied in the literature 
\cite{R7128,R7129,R7130,R7131,R7132,R7133}. More
details about heavy baryons, including the experimental prospects, can be
found in \cite{R7134,R7135}.

Many experimentally measurable quantities such as branching ratio
\cite{R7136}, $\Lambda$ polarization and single-- and double--lepton 
polarizations have already been studied in \cite{R7137,R7138} and 
\cite{R7139}, 
respectively. Analysis of such quantities can be useful for more precise 
determination of the SM parameters and in looking for new physics beyond 
the SM. 

In the present work we analyze the possibility of searching for new physics
in the baryonic $\Lambda_b \rar \Lambda \ell^+ \ell^-$ decay by studying
different asymmetry parameters that characterize the angular dependence of
the angular decay distributions, with the inclusion of 
non--standard vector type of interactions. In our analysis we use the 
helicity amplitude formalism and polarization density matrix method 
(see the first and third references in \cite{R7128}) to analyze the joint
decay distributions in this decay.  

The paper is organized as follows. In section
2, using the Hamiltonian that includes non--standard vector interactions,
the matrix element for the $\Lambda_b \rar \Lambda \ell^+ \ell^-$ is obtained.
In section 3 we calculate the different polarization asymmetries. 
In the final section we study the sensitivity of various asymmetries to the
non--standard interactions.

\section{Matrix element for the $\Lambda_b \rar \Lambda \ell^+ \ell^-$ decay}

In this section we derive the matrix element for the $\Lambda_b \rar \Lambda
\ell^+ \ell^-$ decay which is described by the $b \rar s \ell^+ \ell^-$ 
transition at quark level. Neglecting the terms
proportional to $V_{ub} V_{us}^\ast / V_{tb} V_{ts}^\ast \sim {\cal
O}(10^{-2})$, the matrix element for the $b \rar s \ell^+ \ell^-$
decay can be written in terms of the twelve model independent
four--Fermi interactions as \cite{R7118}

\bea
\label{e7101}
{\cal M} \es \frac{G \alpha}{\sqrt{2} \pi} V_{tb}V_{ts}^\ast \Bigg\{
C_{SL} \bar s_R i \sigma_{\mu\nu} \frac{q^\nu}{q^2} b_L \bar \ell \gamma^\mu
\ell + C_{BR} \bar s_L i \sigma_{\mu\nu} \frac{q^\nu}{q^2} b_R \bar \ell
\gamma^\mu \ell + C_{LL}^{tot} \bar s_L \gamma_\mu b_L \bar \ell_L
\gamma^\mu \ell_L \nnb \\
\ar C_{LR}^{tot} \bar s_L \gamma_\mu b_L \bar \ell_R  
\gamma^\mu \ell_R + C_{RL} \bar s_R \gamma_\mu b_R \bar \ell_L
\gamma^\mu \ell_L + C_{RR} \bar s_R \gamma_\mu b_R \bar \ell_R
\gamma^\mu \ell_R \nnb \\
\ar C_{LRLR} \bar s_L b_R \bar \ell_L \ell_R +
C_{RLLR} \bar s_R b_L \bar \ell_L \ell_R +
C_{LRRL} \bar s_L b_R \bar \ell_R \ell_L +
C_{RLRL} \bar s_R b_L \bar \ell_R \ell_L \nnb \\
\ar C_T \bar s \sigma_{\mu\nu} b \bar \ell \sigma^{\mu\nu} \ell +
i C_{TE} \epsilon_{\mu\nu\alpha\beta} \bar s \sigma^{\mu\nu} b 
\bar \ell \sigma^{\alpha\beta} \ell \Bigg\}~,
\eea
where $q=P_{\Lambda_b} - P_\Lambda = p_1+p_2$ is the momentum transfer and
$C_X$ are the coefficients of the four--Fermi interactions,
$L=(1-\gamma_5)/2$ and $R=(1+\gamma_5)/2$.
The terms with coefficients $C_{SL}$ and
$C_{BR}$ describe the penguin contributions, which correspond to 
$-2 m_s C_7^{eff}$ and $-2 m_b C_7^{eff}$ in the SM, respectively. 
The next four terms in Eq. (\ref{e7101}) with coefficients
$C_{LL}^{tot},~C_{LR}^{tot},~ C_{RL}$ and $C_{RR}$
describe vector type interactions, two ($C_{LL}^{tot}$ and $C_{LR}^{tot}$)
of which contain SM contributions in the form
$C_9^{eff}-C_{10}$ and $C_9^{eff}-C_{10}$, respectively.
Thus, $C_{LL}^{tot}$ and $C_{LR}^{tot}$ can be written as 
\bea
\label{e7102}
C_{LL}^{tot} \es C_9^{eff}- C_{10} + C_{LL}~, \nnb \\
C_{LR}^{tot} \es C_9^{eff}+ C_{10} + C_{LR}~,
\eea
where $C_{LL}$ and $C_{LR}$ describe the contributions of new physics.
Additionally, Eq. (\ref{e7101}) contains four scalar type interactions 
($C_{LRLR},~C_{RLLR},~C_{LRRL}$ and $C_{RLRL}$), and two tensor type 
interactions ($C_T$ and $C_{TE}$). In the present work we will consider the
minimal extension of the SM and therefore we neglect the scalar and
tensor type interactions throughout in this work.

The amplitude of the exclusive $\Lambda_b \rar \Lambda\ell^+ \ell^-$ decay
is obtained by calculating the matrix element of ${\cal H}_{eff}$ for the $b
\rar s \ell^+ \ell^-$ transition between initial and final
baryon states $\lla \Lambda \vel {\cal H}_{eff} \ver \Lambda_b \rra$.
It follows from Eq. (\ref{e7101}) that the matrix elements
\bea
&&\lla \Lambda \vel \bar s \gamma_\mu (1 \mp \gamma_5) b \ver \Lambda_b
\rra~,\nnb \\
&&\lla \Lambda \vel \bar s \sigma_{\mu\nu} (1 \mp \gamma_5) b \ver \Lambda_b
\rra~,\nnb
\eea
are needed in order to calculate
the $\Lambda_b \rar \Lambda\ell^+ \ell^-$ decay amplitude.

These matrix elements parametrized in terms of the form factors are 
as follows (see \cite{R7137,R7140})
\bea
\label{e7103}
\lla \Lambda \vel \bar s \gamma_\mu b \ver \Lambda_b \rra  
\es \bar u_\Lambda \Big[ f_1 \gamma_\mu + i f_2 \sigma_{\mu\nu} q^\nu + f_3  
q_\mu \Big] u_{\Lambda_b}~,\\
\label{e7104}
\lla \Lambda \vel \bar s \gamma_\mu \gamma_5 b \ver \Lambda_b \rra
\es \bar u_\Lambda \Big[ g_1 \gamma_\mu \gamma_5 + i g_2 \sigma_{\mu\nu}
\gamma_5 q^\nu + g_3 q_\mu \gamma_5\Big] u_{\Lambda_b}~, \\
\label{e7105}
\lla \Lambda \vel \bar s \sigma_{\mu\nu} b \ver \Lambda_b \rra
\es \bar u_\Lambda \Big[ f_T \sigma_{\mu\nu} - i f_T^V \ga \gamma_\mu q^\nu -
\gamma_\nu q^\mu \dr - i f_T^S \ga P_\mu q^\nu - P_\nu q^\mu \dr \Big]
u_{\Lambda_b}~,\\
\label{e7106}
\lla \Lambda \vel \bar s \sigma_{\mu\nu} \gamma_5 b \ver \Lambda_b \rra
\es \bar u_\Lambda \Big[ g_T \sigma_{\mu\nu} - i g_T^V \ga \gamma_\mu q^\nu -
\gamma_\nu q^\mu \dr - i g_T^S \ga P_\mu q^\nu - P_\nu q^\mu \dr \Big]
\gamma_5 u_{\Lambda_b}~,
\eea
where $P = p_{\Lambda_b} + p_\Lambda$ and $q= p_{\Lambda_b} - p_\Lambda$. 

The form factors of the magnetic dipole operators are defined as 
\bea
\label{e7107}
\lla \Lambda \vel \bar s i \sigma_{\mu\nu} q^\nu  b \ver \Lambda_b \rra
\es \bar u_\Lambda \Big[ f_1^T \gamma_\mu + i f_2^T \sigma_{\mu\nu} q^\nu
+ f_3^T q_\mu \Big] u_{\Lambda_b}~,\nnb \\
\lla \Lambda \vel \bar s i \sigma_{\mu\nu}\gamma_5  q^\nu  b \ver \Lambda_b \rra
\es \bar u_\Lambda \Big[ g_1^T \gamma_\mu \gamma_5 + i g_2^T \sigma_{\mu\nu}
\gamma_5 q^\nu + g_3^T q_\mu \gamma_5\Big] u_{\Lambda_b}~.
\eea

Using the identity 
\bea
\sigma_{\mu\nu}\gamma_5 = - \frac{i}{2} \epsilon_{\mu\nu\alpha\beta}
\sigma^{\alpha\beta}~,\nnb
\eea
and Eq. (\ref{e7105}), the last expression in Eq. (\ref{e7107}) can be written as
\bea
\lla \Lambda \vel \bar s i \sigma_{\mu\nu}\gamma_5  q^\nu  b \ver \Lambda_b \rra
\es \bar u_\Lambda \Big[ f_T i \sigma_{\mu\nu} \gamma_5 q^\nu \Big]
u_{\Lambda_b}~.\nnb
\eea  
Multiplying (\ref{e7105}) and (\ref{e7106}) by $i q^\nu$ and comparing with
(\ref{e7107}), one can easily obtain the following relations between the form
factors
\bea
\label{e7108}
f_2^T \es f_T + f_T^S q^2~,\crcr
f_1^T \es \Big[ f_T^V + f_T^S \ga m_{\Lambda_b} + m_\Lambda\dr \Big] 
q^2~ = - \frac{q^2}{m_{\Lambda_b} - m_\Lambda} f_3^T~,\nnb \\
g_2^T \es g_T + g_T^S q^2~,\\
g_1^T \es \Big[ g_T^V - g_T^S \ga m_{\Lambda_b} - m_\Lambda\dr \Big]
q^2 =  \frac{q^2}{m_{\Lambda_b} + m_\Lambda} g_3^T~.\nnb
\eea 

Using these definitions of the form factors, for the matrix element
of the $\Lambda_b \rar \Lambda\ell^+ \ell^-$ we get \cite{R7137,R7138}
\bea
\label{e7109}
{\cal M} \es \frac{G \alpha}{4 \sqrt{2}\pi} V_{tb}V_{ts}^\ast \frac{1}{2} \Bigg\{
\bar{\ell} \gamma_\mu (1-\gamma_5) \ell \, 
\bar{u}_\Lambda \Big[ (A_1 - D_1) \gamma_\mu (1+\gamma_5) +
(B_1 - E_1) \gamma_\mu (1-\gamma_5) \nnb \\
\ar i \sigma_{\mu\nu} q^\nu \Big( (A_2 - D_2) (1+\gamma_5) +
(B_2 - E_2) (1-\gamma_5) \Big) \Big] u_{\Lambda_b} \nnb \\
\ar \bar{\ell} \gamma_\mu (1+\gamma_5) \ell \, 
\bar{u}_\Lambda \Big[ (A_1 + D_1) \gamma_\mu (1+\gamma_5) +
(B_1 + E_1) \gamma_\mu (1-\gamma_5) \nnb \\
\ar i \sigma_{\mu\nu} q^\nu \Big( (A_2 + D_2) (1+\gamma_5) +
(B_2 + E_2) (1-\gamma_5) \Big) \nnb \\
\ar q_\mu \Big( (A_3 + D_3) (1+\gamma_5) + (B_3 + D_2) (1-\gamma_5) \Big)
\Big] u_{\Lambda_b} \Bigg\}~,
\eea
where
\bea
\label{e7110}
A_1 \es \frac{1}{q^2}\ga f_1^T-g_1^T \dr C_{SL} + \frac{1}{q^2}\ga
f_1^T+g_1^T \dr C_{BR} + \frac{1}{2}\ga f_1-g_1 \dr \ga C_{LL}^{tot} +
C_{LR}^{tot} \dr \nnb \\
\ar \frac{1}{2}\ga f_1+g_1 \dr \ga C_{RL} + C_{RR} \dr~,\nnb \\
A_2 \es A_1 \ga 1 \rar 2 \dr ~,\nnb \\
A_3 \es A_1 \ga 1 \rar 3 \dr ~,\nnb \\
B_1 \es A_1 \ga g_1 \rar - g_1;~g_1^T \rar - g_1^T \dr ~,\nnb \\
B_2 \es B_1 \ga 1 \rar 2 \dr ~,\nnb \\
B_3 \es B_1 \ga 1 \rar 3 \dr ~,\nnb \\
D_1 \es \frac{1}{2} \ga C_{RR} - C_{RL} \dr \ga f_1+g_1 \dr +
\frac{1}{2} \ga C_{LR}^{tot} - C_{LL}^{tot} \dr \ga f_1-g_1 \dr~,\nnb \\
D_2 \es D_1 \ga 1 \rar 2 \dr ~, \\
D_3 \es D_1 \ga 1 \rar 3 \dr ~,\nnb \\
E_1 \es D_1 \ga g_1 \rar - g_1 \dr ~,\nnb \\
E_2 \es E_1 \ga 1 \rar 2 \dr ~,\nnb \\
E_3 \es E_1 \ga 1 \rar 3 \dr ~.\nnb
\eea

From these expressions it follows
that $\Lambda_b \rar\Lambda \ell^+\ell^-$ decay is described in terms of  
many form factors. It is shown in \cite{R7141} that Heavy Quark Effective
Theory reduces
the number of independent form factors to two ($F_1$ and
$F_2$) irrelevant of the Dirac structure
of the corresponding operators, i.e., 
\bea
\label{e7111}
\lla \Lambda(p_\Lambda) \vel \bar s \Gamma b \ver \Lambda(p_{\Lambda_b})
\rra = \bar u_\Lambda \Big[F_1(q^2) + \not\!v F_2(q^2)\Big] \Gamma
u_{\Lambda_b}~,
\eea
where $\Gamma$ is an arbitrary Dirac structure and
$v^\mu=p_{\Lambda_b}^\mu/m_{\Lambda_b}$ is the four--velocity of
$\Lambda_b$. Comparing the general form of the form factors given in Eqs.
(\ref{e7104})--(\ref{e7108}) with (\ref{e7111}), one can
easily obtain the following relations among them (see also
\cite{R7137,R7138,R7140})
\bea
\label{e7112}
g_1 \es f_1 = f_2^T= g_2^T = F_1 + \sqrt{\hat{r}_\Lambda} F_2~, \nnb \\
g_2 \es f_2 = g_3 = f_3 = g_T^V = f_T^V = \frac{F_2}{m_{\Lambda_b}}~,\nnb \\
g_T^S \es f_T^S = 0 ~,\nnb \\
g_1^T \es f_1^T = \frac{F_2}{m_{\Lambda_b}} q^2~,\nnb \\
g_3^T \es \frac{F_2}{m_{\Lambda_b}} \ga m_{\Lambda_b} + m_\Lambda \dr~,\nnb \\
f_3^T \es - \frac{F_2}{m_{\Lambda_b}} \ga m_{\Lambda_b} - m_\Lambda \dr~,
\eea
where $\hat{r}_\Lambda=m_\Lambda^2/m_{\Lambda_b}^2$.

In order to obtain the helicity amplitudes for the $\Lambda_b \rar\Lambda
\ell^+\ell^-$ decay, it is convenient to regard this decay as a quasi 
two--body decay $\Lambda_b \rar\Lambda V^\ast$ followed by the leptonic
decay $V^\ast \rar  \ell^+\ell^-$, where $V^\ast$ is the off--shell $\gamma$
or $Z$ bosons. The matrix element of $\Lambda_b \rar\Lambda \ell^+\ell^-$
decay can be written in the following form:
\bea
{\cal M}_{\lambda_i}^{\lambda_\ell \bar{\lambda}_\ell} =
\sum_{\lambda_{V^\ast}} \eta_{\lambda_{V^\ast}}
L_{\lambda_{V^\ast}}^{\lambda_\ell \bar{\lambda}_\ell} \, 
H_{\lambda_{V^\ast}}^{\lambda_i}~, \nnb
\eea
where 
\bea
\label{e7113}
L_{\lambda_{V^\ast}}^{\lambda_\ell \bar{\lambda}_\ell} \es 
\varepsilon_{V^\ast}^\mu \lla \ell^-(p_\ell,\lambda_\ell) \,
\ell^+(p_\ell,\bar{\lambda}_\ell) \vel J_\mu^\ell \ver 0 \rra~,\\ 
\label{e7114}
H_{\lambda_{V^\ast}}^{\lambda_i} \es 
\ga \varepsilon_{V^\ast}^\mu \dr^\ast \lla \Lambda
(p_\Lambda,\lambda_\Lambda) \vel J_\mu^i 
\ver \Lambda_b(p_{\Lambda_b}) \rra~,
\eea
where $\varepsilon_{V^\ast}^\mu$ is the polarization vector of the virtual
intermediate vector boson. The metric tensor can be expressed in terms of
the polarization vector of the virtual vector particle
$\varepsilon_V = \varepsilon (\lambda_V)$ as
\bea
-g^{\mu\nu} = \sum_{\lambda_{V^\ast}} \eta_{\lambda_{V^\ast}}
\varepsilon_{\lambda_{V^\ast}}^\mu \varepsilon_{\lambda_{V^\ast}}^{\ast\nu}~,\nnb
\eea  
where the summation is over the helicity of the virtual vector particle
$V,~\Lambda_V = \pm 1, 0, t$ with the metric $\eta_\pm = \eta_0 = - \eta_t =
1$, where $\lambda_V = t$ is the scalar (zero) helicity component of the
virtual $V$ particle (for more details see \cite{R7142,R7143} and first and
third references in \cite{R7128}).
The upper indices in Eqs. (\ref{e7113}) and (\ref{e7114}) correspond to the 
helicities of the leptons and the lower ones correspond to the helicity of 
the $\Lambda$ baryon. Moreover, $J_\mu^\ell$ and $J_\mu^i$ in Eqs. (\ref{e7113}) 
and (\ref{e7114}) are the leptonic and hadronic currents, respectively. 

In the calculations of the leptonic and baryonic amplitudes we will use two
different frames. The leptonic amplitude $L_{\lambda_{V^\ast}}^{\lambda_\ell
\bar{\lambda}_\ell}$ is calculated in the rest frame of the virtual vector
boson wit the z--axis chosen along the $\Lambda$ direction and the x--z
plane chosen as the virtual $V$ decay plane. The hadronic amplitude is
calculated in the rest frame of $\Lambda_b$ baryon.

Using Eqs. (\ref{e7109})--(\ref{e7114}), after lengthy calculations, we get
for the helicity amplitudes:
\bea
\label{e7115}
{\cal M}_{+1/2}^{++} \es 2 m_\ell \sin\theta \Big( H_{+1/2,+1}^{(1)} +
H_{+1/2,+1}^{(2)}\Big) + 2 m_\ell \cos\theta \Big( H_{+1/2,0}^{(1)} +
H_{+1/2,0}^{(2)}\Big) \nnb \\
\ar 2 m_\ell \Big( H_{+1/2,t}^{(1)} -
H_{+1/2,t}^{(2)}\Big)~, \nnb \\
{\cal M}_{+1/2}^{+-} \es - \sqrt{q^2} (1-\cos\theta) \Big[
(1-v) H_{+1/2,+1}^{(1)} + (1+v) H_{+1/2,+1}^{(2)}\Big] -
\sqrt{q^2} \sin\theta  \Big[(1-v) H_{+1/2,0}^{(1)} \nnb \\
\ar (1+v) H_{+1/2,0}^{(2)}\Big]~, \nnb \\
{\cal M}_{+1/2}^{-+} \es \sqrt{q^2} (1+\cos\theta) \Big[
(1+v) H_{+1/2,+1}^{(1)} + (1-v) H_{+1/2,+1}^{(2)}\Big] -
\sqrt{q^2} \sin\theta  \Big[(1+v) H_{+1/2,0}^{(1)} \nnb \\
\ar (1-v) H_{+1/2,0}^{(2)}\Big]~, \nnb \\
{\cal M}_{+1/2}^{--} \es - 2 m_\ell \sin\theta \Big(
H_{+1/2,+1}^{(1)} +
H_{+1/2,+1}^{(2)}\Big) - 2 m_\ell \cos\theta \Big( H_{+1/2,0}^{(1)} +
H_{+1/2,0}^{(2)}\Big) \nnb \\
\ar 2 m_\ell \Big( H_{+1/2,t}^{(1)} -
H_{+1/2,t}^{(2)}\Big)~, \nnb \\
{\cal M}_{-1/2}^{++} \es - 2 m_\ell \sin\theta \Big(
H_{-1/2,-1}^{(1)} +
H_{-1/2,-1}^{(2)}\Big) + 2 m_\ell \cos\theta \Big( H_{-1/2,0}^{(1)} +
H_{-1/2,0}^{(2)}\Big) \nnb \\
\ar 2 m_\ell \Big( H_{-1/2,t}^{(1)} -
H_{-1/2,t}^{(2)}\Big)~, \nnb \\
{\cal M}_{-1/2}^{+-} \es - \sqrt{q^2} (1+\cos\theta) \Big[
(1-v) H_{-1/2,-1}^{(1)} + (1+v) H_{-1/2,-1}^{(2)}\Big] -
\sqrt{q^2} \sin\theta  \Big[(1-v) H_{-1/2,0}^{(1)} \nnb \\
\ar (1+v) H_{-1/2,0}^{(2)}\Big]~, \nnb \\
{\cal M}_{-1/2}^{-+} \es \sqrt{q^2} (1-\cos\theta) \Big[
(1+v) H_{-1/2,-1}^{(1)} + (1-v) H_{-1/2,-1}^{(2)}\Big] -
\sqrt{q^2} \sin\theta  \Big[(1+v) H_{-1/2,0}^{(1)} \nnb \\
\ar (1-v) H_{-1/2,0}^{(2)}\Big]~, \nnb \\
{\cal M}_{-1/2}^{--} \es 2 m_\ell \sin\theta \Big( H_{-1/2,-1}^{(1)} +
H_{-1/2,-1}^{(2)}\Big) - 2 m_\ell \cos\theta \Big( H_{-1/2,0}^{(1)} +
H_{-1/2,0}^{(2)}\Big) \nnb \\
\ar 2 m_\ell \Big( H_{-1/2,t}^{(1)} -
H_{-1/2,t}^{(2)}\Big)~,
\eea
where
\bea      
\label{e7116}
H_{\pm 1/2,\pm1}^{(1)}   \es H_{1/2,1}^{(1)\,V}   \pm H_{1/2,1}^{(1)\,A}~,   \nnb \\
H_{\pm 1/2,\pm1}^{(2)}   \es H_{1/2,1}^{(2)\,V}   \pm H_{1/2,1}^{(2)\,A}~,   \nnb \\
H_{\pm 1/2,0}^{(1,2)}    \es H_{1/2,0}^{(1,2)\,V} \pm H_{1/2,1}^{(1,2)\,A}~, \nnb \\
H_{\pm 1/2,t}^{(1,2)}    \es H_{1/2,t}^{(1,2)\,V} \pm H_{1/2,t}^{(1,2)\,A}~,
\eea
where $\theta$ is the angle of the positron in the rest frame of the
intermediate boson with respect to its helicity axes.
Explicit expressions of the helicity amplitudes $H_{\lambda,\lambda_W}^{V,A}$
are
\bea
\label{e7117}
H_{1/2,1}^{(1)\,V} \es - \sqrt{Q_-} \Big[ F_1^V - (m_{\Lambda_b}+m_\Lambda) F_2^V
\Big]~, \nnb \\
H_{1/2,1}^{(1)\,A} \es - \sqrt{Q_+} \Big[ F_1^A + (m_{\Lambda_b}-m_\Lambda) F_2^A
\Big]~, \nnb \\
H_{1/2,1}^{(2)\,V} \es H_{1/2,1}^{(1)\,V} (F_1^V \rar F_3^V,~F_2^V \rar
F_4^V)~, \nnb \\
H_{1/2,1}^{(2)\,A} \es H_{1/2,1}^{(1)\,A} (F_1^A \rar F_3^A,~F_2^A \rar
F_4^A)~, \nnb \\
H_{1/2,0}^{(1)\,V} \es - \frac{1}{\sqrt{q^2}} \Big\{ \sqrt{Q_-} 
\Big[ (m_{\Lambda_b}+m_\Lambda) F_1^V - q^2 F_2^V \Big]
\Big\}~,\nnb \\
H_{1/2,0}^{(1)\,A} \es - \frac{1}{\sqrt{q^2}} \Big\{ \sqrt{Q_+} 
\Big[ (m_{\Lambda_b}-m_\Lambda) F_1^A + q^2 F_2^A \Big]
\Big\}~,\nnb \\
H_{1/2,0}^{(2)\,V} \es H_{1/2,0}^{(1)\,V} (F_1^V \rar F_3^V,~F_2^V \rar
F_4^V) ~,\nnb \\
H_{1/2,0}^{(2)\,A} \es H_{1/2,0}^{(1)\,A} (F_1^A \rar F_3^A,~F_2^A \rar
F_4^A) ~,\nnb \\
H_{1/2,t}^{(1)\,V} \es - \frac{1}{\sqrt{q^2}} \Big\{ \sqrt{Q_+} 
\Big[ (m_{\Lambda_b}-m_\Lambda) F_1^V + q^2 F_5^V \Big]
\Big\}~,\nnb \\
H_{1/2,t}^{(1)\,A} \es - \frac{1}{\sqrt{q^2}} \Big\{ \sqrt{Q_-} 
\Big[ (m_{\Lambda_b}+m_\Lambda) F_1^A - q^2 F_5^A \Big]
\Big\}~,\nnb \\
H_{1/2,t}^{(2)\,V} \es H_{1/2,t}^{(1)\,V} (F_1^V \rar F_3^V,~F_5^V \rar 
F_6^V) ~,\nnb \\
H_{1/2,t}^{(2)\,A} \es H_{1/2,t}^{(1)\,A} (F_1^A \rar F_3^A,~F_5^A \rar
F_6^A) ~,
\eea
where
\bea
Q_+ \es (m_{\Lambda_b}+m_\Lambda)^2 - q^2~,\nnb \\
Q_- \es (m_{\Lambda_b}-m_\Lambda)^2 - q^2~,\nnb
\eea 
and
\bea
\label{e7118}
F_1^V \es  A_1-D_1+B_1-E_1~,\nnb \\
F_1^A \es  A_1-D_1-B_1+E_1~,\nnb \\
F_2^V \es  F_1^V (1\rar 2)~,\nnb \\
F_2^A \es  F_1^A (1\rar 2)~,\nnb \\
F_3^V \es  A_1+D_1+B_1+E_1~,\nnb \\
F_3^A \es  A_1+D_1-B_1-E_1~,\nnb \\
F_4^V \es  F_3^V (1\rar 2)~,\nnb \\
F_4^A \es  F_3^A (1\rar 2)~,\nnb \\
F_5^V \es  F_1^V (1\rar 3)~,\nnb \\
F_5^A \es  F_1^A (1\rar 3)~,\nnb \\
F_6^V \es  F_3^V (1\rar 3)~,\nnb \\
F_6^A \es  F_3^A (1\rar 3)~.
\eea
The remaining helicity amplitudes can be obtained from the parity relations
\bea
\label{e7119}
H_{-\lambda,-\lambda_W}^{V,(A)} = +(-) H_{\lambda,\lambda_W}^{V,(A)}~.
\eea
The square of the matrix element for the $\Lambda_b \rar\Lambda \ell^+\ell^-$ 
decay is given as
\bea
\label{e7120}
\vel {\cal M} \ver^2 \es \vel {\cal M}_{+1/2}^{++} \ver^2 +
\vel {\cal M}_{+1/2}^{+-} \ver^2 + \vel {\cal M}_{+1/2}^{-+} \ver^2 + 
\vel {\cal M}_{+1/2}^{--} \ver^2 \nnb \\
\ar \vel {\cal M}_{-1/2}^{++} \ver^2 +
\vel {\cal M}_{-1/2}^{+-} \ver^2 + \vel {\cal M}_{-1/2}^{-+} \ver^2
+ \vel {\cal M}_{-1/2}^{--} \ver^2~.
\eea
Following the standard methods used in literature (see the third
reference in \cite{R7128}), the normalized joint angular decay
distribution for the two cascade decay
\bea
\Lambda_b^{1/2 ^+} \rar \Lambda^{1/2 ^+} \Big( \rar a(1/2 ^+) +
b(0^-)\Big)
+ V (\rar \ell^+ \ell^-)~,\nnb 
\eea
 can be written as
\bea
\label{e7121}
\frac{\ds d \Gamma}{\ds dq^2 d\!\cos\theta \,d\!\cos\theta_\Lambda} \es 
\vel \frac{G \alpha}{4 \sqrt{2} \pi} V_{tb} V_{ts}^\ast \frac{1}{2} \ver^2 
\frac{\sqrt{\lambda(m_{\Lambda_b}^2,m_\Lambda^2,q^2)}
\sqrt{\lambda(m_\Lambda^2,m_a^2,m_b^2)}}
{1024 \pi^3 m_{\Lambda_b}^3 m_\Lambda^2} 
 v {\cal B}(\Lambda_b \rar a + b) \vel {\cal M} \ver^2~, \nnb \\
\eea
where the polar angle $\theta_\Lambda$ is the angle of the $a(1/2 ^+)$
momentum in the rest frame of the $\Lambda$ baryon. Note that in this
expression we perform integration over the azimuthal angle $\varphi$
between the planes of the two decays $\Lambda \rar a + b$ and 
$V \rar \ell^+ \ell^-$. Our final result for the differential decay width is  
\bea
\label{e7122}
\lefteqn{
\frac{\ds d \Gamma}{\ds dq^2 d\!\cos\theta \,d\!\cos\theta_\Lambda} =
\vel \frac{G \alpha}{4 \sqrt{2} \pi} V_{tb} V_{ts}^\ast \frac{1}{2} \ver^2
\frac{\sqrt{\lambda(m_{\Lambda_b}^2,m_\Lambda^2,q^2)}
\sqrt{\lambda(m_\Lambda^2,m_a^2,m_b^2)}}{1024 \pi^3 m_{\Lambda_b}^3 m_\Lambda^2}
v {\cal B}(\Lambda \rar a + b) }\nnb \\ 
&&\Bigg\{
(1+\alpha_\Lambda \cos \theta_\Lambda) \Bigg[ \Big(8 m_\ell^2 \sin^2\theta 
\vel A_{+1/2,+1} \ver^2 + (1-\cos\theta)^2 q^2 
\vel A_{+1/2,+1}- v B_{+1/2,+1} \ver^2 \nnb \\
&&+ (1+\cos\theta)^2 q^2 
\vel A_{+1/2,+1}+ v B_{+1/2,+1} \ver^2 \Big) +
 8 m_\ell^2 \cos^2\theta \vel A_{+1/2,0} \ver^2 +
8 m_\ell^2 \vel B_{+1/2,t} \ver^2 \nnb \\
&& + \sin^2\theta q^2 \Big( 2 \vel A_{+1/2,0} \ver^2 + 2 v^2 \vel B_{+1/2,0} \ver^2
\Big)\Bigg] \nnb \\
&& + (1-\alpha_\Lambda \cos \theta_\Lambda) \Bigg[ \Big(8 m_\ell^2 \sin^2\theta 
\vel A_{-1/2,-1} \ver^2 + (1+\cos\theta) q^2 
\vel A_{-1/2,-1}- v B_{-1/2,-1} \ver^2 \nnb \\
&&+ (1-\cos\theta)^2 q^2 
\vel A_{-1/2,-1}+ v B_{-1/2,-1} \ver^2 \Big) +
8 m_\ell^2 \cos^2\theta \vel A_{-1/2,0} \ver^2 +
8 m_\ell^2 \vel B_{-1/2,t} \ver^2 \nnb \\
&& + \sin^2\theta q^2 \Big( 2 \vel A_{-1/2,0} \ver^2 + 2 v^2 \vel B_{-1/2,0} \ver^2
\Big)\Bigg]\Bigg\}~.
\eea

In Eq. (\ref{e7122}) we induce the following definitions:
\bea
\label{e7123}
H_{\lambda_i,\lambda_W}^{(1)} + H_{\lambda_i,\lambda_W}^{(2)} \es 
A_{\lambda_i,\lambda_W}\nnb \\
H_{\lambda_i,\lambda_W}^{(1)} - H_{\lambda_i,\lambda_W}^{(2)} \es 
B_{\lambda_i,\lambda_W}~.
\eea

One can easily see that in addition to the variables that exist in
Eq. (\ref{e7122}), there appears a new variable $\theta_\Lambda$ and 
integration of Eq. (\ref{e7122}) over it gives  the differential decay 
width for the $\Lambda_b \rar \Lambda \ell^+\ell^-$ decay. 

It is well known
that heavy quarks $b(c)$ resulting from $Z$ decay are polarized. It is shown
in \cite{R7144,R7145} that a sizeable fraction of the $b$ quark polarization
retained in fragmentation of heavy quarks to heavy baryons. Therefore, an
additional set of polarization observables can be obtained if the
polarization of the heavy $\Lambda_b$ baryon is taken into account.

In order to take polarization of the $\Lambda_b$ baryon into consideration    
we will use the density matrix method. The spin density matrix of $\Lambda$
baryon is
\bea
\label{e7124}
\rho = \frac{1}{2}
\left(
\begin{array}{cc}
1+{\cal P}\cos\theta_\Lambda^S & {\cal P}\cos\theta_\Lambda^S \\ \\
{\cal P}\cos\theta_\Lambda^S   & 1-{\cal P}\cos\theta_\Lambda^S
\end{array}
\right)~,
\eea
where ${\cal P}$ is the polarization of $\Lambda_b$, and $\theta_\Lambda^S$
is the angle that the polarization of $\Lambda_b$ makes with the momentum of
$\Lambda$, in the rest frame of $\Lambda_b$.  

The four--fold decay distribution can easily be obtained from Eq.
(\ref{e7122}). Obviously, there appears on the left--hand side of Eq.
(\ref{e7122}) the distribution over $\theta_\Lambda^S$, i.e., 
$d/d\cos\theta_\Lambda^S$. Hence the right--hand side of the same equation
can be modified as follows:
\bea
\label{e7125}
&&\vel +1/2,+1 \ver^2 \rar (1-{\cal P}\cos\theta_\Lambda^S) 
\vel +1/2,+1 \ver^2~, \nnb \\
&&\Big\{\vel +1/2,t \ver^2,~\vel +1/2,0 \ver^2,~ (+1/2,t) (+1/2,0)^\ast\Big\} 
\rar (1+{\cal P}\cos\theta_\Lambda^S)
\Big\{ \vel +1/2,t \ver^2,\vel +1/2,0 \ver^2,\nnb \\
&&(+1/2,t) (+1/2,0)^\ast \Big\}~,\nnb \\ 
&&\Big\{(+1/2,+1) (+1/2,t)^\ast,~(+1/2,+1) (+1/2,0)^\ast \Big\} \rar
{\cal P}\sin\theta_\Lambda^S \Big\{(+1/2,+1) (+1/2,t)^\ast,\nnb \\
&&(+1/2,1) (+1/2,0)^\ast \Big\}~, \nnb \\
&&\vel -1/2,-1 \ver^2 \rar (1+{\cal P}\cos\theta_\Lambda^S) \vel -1/2,-1
\ver^2~, \nnb \\
&&\Big\{\vel -1/2,t \ver^2,~\vel - 1/2,0 \ver^2,~(-1/2,t)
(-1/2,0)^\ast\Big\} 
\rar (1-{\cal P}\cos\theta_\Lambda^S)
\Big\{ \vel -1/2,t \ver^2,~\vel -1/2,0 \ver^2,\nnb \\
&&(-1/2,t) (-1/2,0)^\ast \Big\}~,\nnb \\
&&\Big\{(-1/2,-1) (-1/2,t)^\ast,~(-1/2,-1) (-1/2,0)^\ast \Big\} \rar
{\cal P}\sin\theta_\Lambda^S \Big\{(-1/2,-1) (-1/2,t)^\ast,\nnb \\
&&(-1/2,-1)(-1/2,0)^\ast \Big\}~.
\eea

It follows from Eqs. (\ref{e7122}) and (\ref{e7124}) that the cascade decay
$\Lambda_b \rar \Lambda(\rar a+b) \, V^\ast(\rar \ell^+ \ell^-)$ has a rich angular
structure. Therefore, study of different distributions will prove useful in
separating various angular coefficients in the experiments. For this reason,
instead of analyzing the full four--fold angular distributions, one can
investigate the individual angular  distributions and their relations to
the new Wilson coefficients. Foe example, the polar angle $\theta_\Lambda$
distribution of the cascade decay $\Lambda \rar a+b$ can be obtained from
Eq. (\ref{e7122}) by performing integration over $\theta$, as a result of
which takes the form
\bea
\label{e7126}
\frac{d\Gamma}{dq^2\,d\cos\theta_\Lambda} \sim 1 + \alpha \alpha_\Lambda
\cos\theta_\Lambda~,
\eea
where the asymmetry parameter $\alpha$ is defined as
\bea
\label{e7127}
\alpha \es \frac{8}{3 \Delta}\Big\{4 m_\ell^2 \vel A_{+1/2,+1} \ver^2 + 2 q^2 
\Big(\vel A_{+1/2,+1} \ver^2 + v^2 \vel B_{+1/2,+1} \ver^2
\Big) \nnb \\
\ar 2 m_\ell^2 \vel A_{+1/2,0} \ver^2
+ 6 m_\ell^2 \vel B_{+1/2,t} \ver^2 + q^2 \Big(\vel A_{+1/2,0} 
\ver^2 + v^2 \vel B_{+1/2,0} \ver^2 \Big) \nnb \\
\ek 4 m_\ell^2 \vel A_{-1/2,-1} \ver^2 - 2 q^2 
\Big(\vel A_{-1/2,-1} \ver^2 + v^2 \vel B_{-1/2,-1} \ver^2 \Big) \nnb \\
\ek q^2 \big(\vel A_{-1/2,0} \ver^2 + v^2 \vel B_{-1/2,0} \ver^2\Big)
- 2 m_\ell^2 \vel A_{-1/2,0} \ver^2 - 6 m_\ell^2 
\vel B_{-1/2,t} \ver^2 \Big\}~,
\eea
where
\bea
\label{e7128}
\Delta \es \frac{8}{3}\Big\{4 m_\ell^2 \vel A_{+1/2,+1} \ver^2 + 2 q^2 
\Big(\vel A_{+1/2,+1} \ver^2 + v^2 \vel B_{+1/2,+1} \ver^2
\Big) \nnb \\
\ar 2 m_\ell^2 \vel A_{+1/2,0} \ver^2
+ 6 m_\ell^2 \vel B_{+1/2,t} \ver^2 + q^2 \Big(\vel A_{+1/2,0} 
\ver^2 + v^2 \vel B_{+1/2,0} \ver^2 \Big) \nnb \\
\ar 4 m_\ell^2 \vel
A_{-1/2,-1} \ver^2 + 2 q^2 
\Big(\vel A_{-1/2,-1} \ver^2 + v^2 \vel B_{-1/2,-1} \ver^2 \Big) \nnb \\
\ar q^2 \big(\vel A_{-1/2,0} \ver^2 + v^2 \vel B_{-1/2,0} \ver^2\Big)
+ 2 m_\ell^2 \vel A_{-1/2,0} \ver^2 + 6 m_\ell^2 
\vel B_{-1/2,t} \ver^2 \Big\}~.
\eea

For the polar angle distribution in the cascade decay $V^\ast \rar \ell^+
\ell^-$ we integrate Eq. (\ref{e7122}) over $\theta_\Lambda$ and we get
\bea
\label{e7129}
\frac{d\Gamma}{dq^2\,d\cos\theta} \sim 1 + 2 \alpha_\theta \cos\theta +
\beta_\theta \cos^2\theta~,
\eea
where
\bea
\label{e7130}
\alpha_\theta \es \frac{1}{\Delta_1} 2 v q^2 \mbox{\rm Re} \Big[
A_{+1/2,+1} B_{+1/2,+1}^\ast - 
A_{-1/2,-1} B_{-1/2,-1}^\ast \Big]~, \\ \nnb \\
\label{e7131}
\beta_\theta \es \frac{1}{\Delta_1}
\Big\{ - 4 m_\ell^2 \vel A_{+1/2,+1} \ver^2 + 
q^2 \Big( \vel A_{+1/2,+1} \ver^2 + v^2 \vel B_{+1/2,+1} \ver^2 \Big) \nnb \\
\ar 4 m_\ell^2 \vel A_{+1/2,0} \ver^2 -
q^2 \Big( \vel A_{+1/2,0} \ver^2 + v^2 \vel B_{+1/2,0} \ver^2 \Big) \nnb \\
\ek 4 m_\ell^2 \vel A_{-1/2,-1} \ver^2 + q^2
\Big( \vel A_{-1/2,-1} \ver^2 + v^2 \vel B_{-1/2,-1} \ver^2 \Big) \nnb \\
\ar 4 m_\ell^2 \vel A_{-1/2,0} \ver^2 -
q^2 \Big( \vel A_{-1/2,0} \ver^2 + v^2 \vel B_{-1/2,0} \ver^2 \Big)
\Big\}~,
\eea
and
\bea
\label{e7132}
\Delta_1 \es 4 m_\ell^2  \vel A_{+1/2,+1} \ver^2 +
q^2 \Big( \vel A_{+1/2,+1} \ver^2 + v^2 \vel B_{+1/2,+1} \ver^2 \Big) \nnb \\
\ar 4 m_\ell^2 \vel B_{+1/2,t} \ver^2
+ q^2 \Big( \vel A_{+1/2,0} \ver^2 + v^2 \vel B_{+1/2,0} \ver^2 \Big) \nnb \\
\ar 4 m_\ell^2 \vel A_{-1/2,-1} \ver^2 + q^2
\Big( \vel A_{-1/2,-1} \ver^2 + v^2 \vel B_{-1/2,-1} \ver^2 \Big) \nnb \\
\ar  4 m_\ell^2 \vel B_{-1/2,t} \ver^2 +
q^2 \Big( \vel A_{-1/2,0} \ver^2 + v^2 \vel B_{-1/2,0} \ver^2 \Big)~,
\eea

If the polarization of the initial $\Lambda_b$ is considered, a new symmetry
parameter, which depends on $\theta_\Lambda^S$ appears. Performing
integrations over $\theta_\Lambda$ and $\theta$, we get
\bea
\label{e7133}
\frac{\ds d\Gamma}{\ds dq^2\,d\!\cos\theta_\Lambda^S} \sim 1 -
\alpha_{\Lambda_S} {\cal P} \cos\theta_\Lambda^S~,
\eea
where
\bea 
\label{e7134}
\alpha_{\Lambda_S} \es \frac{8}{3 \Delta} \Big\{ 4 m_\ell^2 \vel A_{+1/2,+1} \ver^2 +
2 q^2 \Big( \vel A_{+1/2,+1} \ver^2 + v^2 \vel B_{+1/2,+1} \ver^2
\Big) \nnb \\
\ek 2 m_\ell^2 \vel A_{+1/2,0} \ver^2
- q^2 \Big( \vel A_{+1/2,0} \ver^2 + v^2 \vel B_{+1/2,0} \ver^2
\Big)
- 6 m_\ell^2 \vel B_{+1/2,t} \ver^2 \nnb \\
\ek 4 m_\ell^2 \vel A_{-1/2,-1} \ver^2
- 2 q^2 \Big( \vel A_{-1/2,-1} \ver^2 + v^2 \vel B_{-1/2,-1}
\ver^2\Big) \nnb \\
\ar 2 m_\ell^2 \vel A_{-1/2,0} \ver^2+
q^2 \Big( \vel A_{-1/2,0} \ver^2 + v^2 \vel B_{-1/2,0} \ver^2 \Big)
+ 6 m_\ell^2 \vel B_{-1/2,t}
\ver^2\Bigg\}~,
\eea
where $\Delta$ is given in Eq. (\ref{e7128}).

\section{Numerical analysis}

In this section we present our numerical results for the 
asymmetry parameters $\alpha_\theta$, $\alpha_{\theta_\Lambda}$,
$\alpha_{\theta_\Lambda^S}$ and $\beta$. 
The values of the input parameters we use in our
calculations are: $\vel V_{tb} V_{ts}^\ast \ver = 0.0385$, 
$m_\tau = 1.77~GeV$, $m_\mu = 0.106~GeV$.
$m_b = 4.8~GeV$. For the Wilson coefficients we use their SM 
values which are given as: $C_7^{SM} = -0.313$, $C_9^{SM} = 4.344$ and 
$C_{10}^{SM} = -4.669$. In further numerical analysis, the values of the new
Wilson coefficients which describe new physics beyond the SM are needed.
The Wilson coefficients $C_{BR}$ and $C_{SL}$ are strictly constrained from
$b \rar s \gamma$ decay. The SM prediction on the branching
ratio for the $b \rar s \gamma$ decay coincide, practically, with 
experimental result and there seems to be no noticeable deviation 
between them. Therefore we can fix the values of $C_{BR}$ and $C_{SL}$
by substituting their SM values, i.e., $C_{BR}=-2 m_b C_7^{eff}$, 
$C_{SL}=-2 m_s C_7^{eff}$, where $C_7^{eff}=-0.313$. Furthermore some of the
Wilson coefficients describing vector interactions are restricted strongly by
the present experimental data on branching ratios for the $B \rar K \ell^+
\ell^-$ and $B \rar K^\ast \ell^+ \ell^-$ decays \cite{R7125,R7126}. Using
the experimental result on branching ratios for the above--mentioned
decays, we obtain the following restrictions on $C_{LL}$ and $C_{RL}$:
$-2 \le C_{LL} \le 0$, and $0 \le C_{RL} \le 2.3$. The 
remaining Wilson are all varied in the region $-\vel C_{10}^{SM} \ver \le
C_X \le + \vel C_{10}^{SM} \ver$. The upper bound on branching ratio of
$B_s \rar \mu^+ \mu^-$ \cite{R7146} suggests that this is the right order of
magnitude for the vector interaction coefficients. 

Few words about the Wilson coefficients $C_9^{eff}$ are in order. Note
that ${\cal M}(b \rar s \ell^+ \ell^-)$ for the $b \rar s
\ell^+ \ell^-$ decay, although being a free quark decay amplitude, contains
certain long distance effects from matrix elements of the four quark
operators $\lla \ell^+ \ell^- s \vel {\cal O}_i \ver b \rra$ (explicit form
of the operators ${\cal O}_1$--${\cal O}_6$ can be found in
\cite{R7147,R7148}) which are usually combined with the coefficient $C_9$ in an
"effective" Wilson coefficient. For this reason, in exclusive decays one can
define $C_9^{eff}$ as
\bea
\label{e7135}
C_9^{eff}(m_b,\hat s) = C_9(m_b)\left[1 + \frac{\alpha_s(\mu)}{\pi} \omega
(\hat s) \right] + Y_{SD}(m_b,\hat s)~,
\eea
where $Y_{SD}$ corresponds to the above--mentioned four quark operator
matrix elements, and $w (\hat{s})$ represents $\cal{O}(\alpha_s)$
corrections coming from one--gluon exchange in the matrix elements of the
corresponding operator, whose explicit form can be found in \cite{R7147}.
The perturbative calculation leads to the following result for 
$Y_{SD}(\hat{s},m_b)$:
\bea 
Y_{SD}(m_b,\hat{s}) \es g \ga \hat m_c,\hat s \dr
C^{(0)}
- \frac{1}{2} g \ga 1,\hat s \dr
\left[4 C_3 +4 C_4 + 3 C_5 + C_6 \right] \nnb \\
\ek \frac{1}{2} g \ga 0,\hat s \dr
\left[ C_3 + 3  C_4 \right]
+ \frac{2}{9} \left[ 3 C_3 + C_4 + 3 C_5 + C_6 \right]~,\nnb
\eea
where 
\bea
C^{(0)} &=& 3 C_1 + C_2 + 3 C_3 + C_4 + 3 C_5 + C_6~,\nnb
\eea
and the function $g(m_q,s)$ stands for the loops of quarks with
mass $m_q$ at the dilepton invariant mass $s$. This function develops
absorptive parts for dilepton energies $s=4 m_q^2$:
\bea
\lefteqn{
g \ga \hat m_q,\hat s \dr = - \frac{8}{9} \ln \hat m_q +
\frac{8}{27} + \frac{4}{9} y_q -
\frac{2}{9} \ga 2 + y_q \dr \sqrt{\vel 1 - y_q \ver}} \nnb \\
&&\times \Bigg[ \Theta(1 - y_q)
\ga \ln \frac{1  + \sqrt{1 - y_q}}{1  -  \sqrt{1 - y_q}} - i \pi \dr
+ \Theta(y_q - 1) \, 2 \, \arctan \frac{1}{\sqrt{y_q - 1}} \Bigg], \nnb
\eea
where  $\hat m_q= m_{q}/m_{b}$ and $y_q=4 \hat m_q^2/\hat s$

In addition to these perturbative contributions $C_9$ also receives long
distance contributions coming from the production of $\bar{c}c$ resonances
at intermediate states. Their contributions are represented by $Y_{LD}$,
which has the form:
\bea
\label{e7136}
Y_{LD}(\hat s) &=& \frac{3}{\alpha^2} C^{(0)}
\sum_{V_i = \psi \ga 1 s \dr, \cdots, \psi \ga 6 s \dr}
\ds{\frac{ \pi \kappa_{i} \Gamma \ga V_i \rar \ell^+ \ell^- \dr M_{V_i} }
{\ga M_{V_i}^2 - \hat s m_b^2 - i M_{V_i} \Gamma_{V_i} \dr }}~,
\eea
where $\kappa_i$ are the Fudge factors (see for example \cite{R7107}).
In regard to the absorptive parts that $C_9^{eff}$ develops, no new
fermions are introduced, and hence no new sources for the    
additional absorptive parts in the Wilson coefficients occur. 
For this reason we will assume that all new Wilson coefficients are real.

From the expressions of asymmetries it follows that the form
factors are the main and the most important input parameters necessary in
the numerical calculations. The calculation of the form factors of $\Lambda_b
\rar \Lambda$ transition does not exist at present.
But, we can use the results from QCD sum rules
in corporation with HQET \cite{R7141,R7149}. We noted earlier that,
HQET allows us to establish relations among the form factors and reduces
the number of independent form factors into two. 
In \cite{R7141,R7149}, the $q^2$ dependence of these form factors
are given as follows
\bea
F(\hat{s}) = \frac{F(0)}{\ds 1-a_F \hat{s} + b_F \hat{s}^2}~. \nnb
\eea
The values of the parameters $F(0),~a_F$ and $b_F$ are given in table 1.
\begin{table}[h]    
\renewcommand{\arraystretch}{1.5}
\addtolength{\arraycolsep}{3pt}
$$
\begin{array}{|l|ccc|}  
\hline
& F(0) & a_F & b_F \\ \hline
F_1 &
\phantom{-}0.462 & -0.0182 & -0.000176 \\
F_2 &
-0.077 & -0.0685 &\phantom{-}0.00146 \\ \hline
\end{array}
$$
\caption{Form factors for $\Lambda_b \rar \Lambda \ell^+ \ell^-$
decay in a three parameter fit.}
\renewcommand{\arraystretch}{1}
\addtolength{\arraycolsep}{-3pt}
\end{table}  

In order to have an idea about the sensitivity of our results to the
specific parametrization of the two form factors predicted by the QCD sum
rules in corporation with the HQET, we also have used another
parametrization of the form factors based on the pole model and compared the
results of both models. The dipole form of the form factors predicted by the
pole model are given as:
\bea
F_{1,2}(E_\Lambda) = N_{1,2} \left( \frac{\Lambda_{QCD}}{\Lambda_{QCD}+
E_\Lambda} \right)^2~,\nnb
\eea
where
\bea
E_\Lambda = \frac{m_{\Lambda_b}^2 - m_\Lambda^2 - q^2}
{2 m_{\Lambda_b}}~,\nnb
\eea
and $\Lambda_{QCD} = 0.2$, $\vel N_1\ver = 52.32$ and $\vel N_1\ver \simeq
-0.25 N_1$ \cite{R7150}.

From the explicit expressions of the asymmetry parameters we see that they
depend on the new Wilson coefficients and $q^2$. Therefore there might
appear some difficulty in studying the dependence of the physical quantities
on both variables in the experiments. For this reason we will study the
dependence of the asymmetry parameters on $q^2$ at fixed values of the new
Wilson coefficients.

In Fig. (1) we present the dependence of $\alpha$ on $q^2$ for the
$\Lambda_b \rar \Lambda \mu^+ \mu^-$ decay at five fixed values of $C_{RR}$.
We observe from this figure that at all values of $C_{RR}$ the magnitude of
$\alpha$ is smaller compared to the SM case for the whole range of $q^2$.
The dependence of $\alpha$ on $q^2$ is not presented for the Wilson
coefficients $C_{LL}$ and $C_{LR}$, since our numerical analysis yields 
that $\alpha$ is not sensitive the presence of $C_{LL}$ and $C_{LR}$, 
and it coincides with the SM result at all values of $q^2$.

In Fig. (2) we depict the dependence of $\alpha$ on $q^2$ for the $\Lambda_b
\rar \Lambda \mu^+ \mu^-$ decay, at fixed values of $C_{RL}$. From this 
figure we see that, up to $q^2 = 18~GeV^2$ the magnitude of $\alpha$ is
smaller compared to the SM prediction at $C_{RL}=2$, but for $q^2 >
18~GeV^2$ the contribution of $C_{RL}=2$ is exceeds that of the contribution
of $C_{RL}=0$ (i.e., SM case). In other words, investigation of $\alpha$ on 
$q^2$ in different kinematical regions of $q^2$ can give valuable information 
not only about the existence of the new physics, but also about the sign of 
the new Wilson coefficient $C_{RL}$.

The study of the dependence of $\alpha$ on $q^2$ at fixed values of the new
Wilson coefficients for the $\Lambda_b \rar \Lambda \tau^+ \tau^-$ decay leads
to the following results:

\begin{itemize}

\item The dependence of the asymmetry parameter $\alpha$ on $q^2$ is not
sensitive to the presence of $C_{LL}$ and $C_{LR}$, and practically there
seems to be no departure from the SM prediction.

\item The situation drastically changes in the presence of Wilson
coefficients $C_{RL}$ and $C_{RR}$. When $C_{RR}$ ($C_{RL})= 4(2)$, up to the
range $q^2=18~GeV^2$, the value of $\alpha$ is two times smaller (as modulo)
compared to that SM prediction; and when $C_{RR}$ ($C_{RL})=-4(2)$
the departure from from the SM result is about 50\% (30\%) . Therefore 
measurement of the asymmetry parameter $\alpha$ at different values of 
$q^2$ can give useful hint about the existence of $C_{RL}$ and $C_{RR}$.

\end{itemize}  

Next, we analyze the dependence of $\alpha_\theta$ and $\beta_\theta$ 
for the $\Lambda_b \rar \Lambda \mu^+ \mu^-$ decay. Our results can be
summarized as follows:

\begin{itemize}

\item The zero of position of $\alpha_\theta$ is shifted to the right (left)
(see Figs. (3) and (4)) when $C_{LL}$ is negative ($C_{LR}$ is positive).
The essential point here is that, similar to the $B \rar K^\ast \ell^+
\ell^-$ decay, the zero of position of $\alpha_\theta$ is independent of the
long distance effects and determined solely by short distance dynamics only.

\item The zero of position of $\alpha_\theta$ is practically independent of
$C_{RR}$ and $C_{RL}$.

\end{itemize}   

Therefore, determination of zero position of $\alpha_\theta$ 
can serve as a good tool for establishing the new physics beyond the SM, as
well as the sign of new Wilson coefficients, which is controlled by the
short distance physics only.

Moreover, the present analysis shows that $\beta_\theta$ is sensitive to
the existence of the vector interactions $C_{LL}$ in the region $1~GeV^2 \le
q^2 \le 3~GeV^2$, and $C_{LR}$ in the region $1~GeV^2 \le q^2 \le 8~GeV^2$. 
Therefore an 
investigation on the asymmetry parameter $\beta_\theta$ can give useful 
information about the existence of the vector interaction realized by 
the Wilson coefficients $C_{LL}$ and $C_{LR}$. $\beta_\theta$ is not
sensitive to the remaining two vector interactions $C_{RL}$ and $C_{RR}$
for the $\Lambda_b \rar \Lambda \mu^+ \mu^-$ decay.      

From the analysis of the dependence of $\alpha_\theta$ and $\beta_\theta$
on $q^2$ for the $\Lambda_b \rar \Lambda \tau^+ \tau^-$ decay we get:

\begin{itemize}

\item $\alpha_\theta$ shows strong dependence on all Wilson
coefficients.

\item The dependence of $\beta_\theta$ on $q^2$ is similar to the SM case
and at all values of all new Wilson coefficients the sign of $\beta_\theta$
is the same as in the SM case. Far from the resonance regions, it is 
strongly dependent on $C_{LR}$. For example, at
$C_{LR}= \pm 4$, the departure from the SM result is about $50\%$
larger when $14~GeV^2 \le q^2 \le 16~GeV^2$ (see Fig. (5)).

\item At positive (negative) values of $C_{LR}$, the magnitude of 
$\beta_\theta$ is smaller (larger) compared to that of SM prediction.

\end{itemize} 

Finally, let us discuss the dependence of the asymmetry parameter
$\alpha_{\Lambda_S}$ on $q^2$ at fixed values of $C_X$. In the
$\Lambda_b \rar \Lambda \mu^+ \mu^-$ decay, $\alpha_{\Lambda_S}$ is more 
sensitive to all vector interactions in the kinematical region $1~GeV^2 \le q^2
\le 5~GeV^2$ (see Figs. (6) and (7), respectively).

For the $\Lambda_b \rar \Lambda \tau^+ \tau^-$ decay
$\alpha_{\Lambda_S}$ is  sensitive to all type of vector interactions
(see Figs. (8)--(11)), and it exhibits different behavior in its dependence
on the new Wilson coefficients. 

The dependence of $\alpha_{\Lambda_S}$ on the Wilson coefficients $C_{LR}$ is 
similar to its dependence on $C_{LL}$. In the same region of $q^2$ when
$C_{LR}=\pm 4$, $\alpha_{\Lambda_S}$ is two times smaller compared to that of 
the SM result.  
Note that near the end of the spectrum, i.e., $17.6~GeV^2 \le q^2 \le
19.6~GeV^2$, $\alpha_{\Lambda_S}$ changes its sign when $C_{LR}=+4$ (see
Fig. (9)). For all other type of vector interactions, the asymmetry
parameter $\alpha_{\Lambda_S}$ does not change its sign.

Therefore determination of the values of $\alpha_{\Lambda_S}$ in experiments
can serve as an efficient tool for establishing the existence of the new 
type of vector interactions and also their signs. 

In conclusion, in the present work we calculate the helicity amplitudes in
the $\Lambda_b \rar \Lambda \ell^+ \ell^-$ decay in the framework of the
minimal extension of the standard model with the inclusion of the new vector
interactions. We analyze various asymmetry parameters of the 
$\Lambda_b \rar \Lambda(\rar a+b) \, V^\ast (\rar \ell^+ \ell^-)$
decay with polarized and unpolarized heavy baryons and study their dependence 
on $q^2$ at fixed values of the new vector type interaction Wilson coefficients. 
We considered different asymmetry parameters and obtain that they exhibit strong
dependence on different new Wilson coefficients. Therefore measurement of
the different asymmetry parameters, namely, $\alpha$, $\alpha_\theta$,
$\beta_\theta$ and $\alpha_{\Lambda_S}$, can give conformative informative
about the existence of the new physics beyond the SM. 
\newpage
 
\section*{Acknowledgments}

One of the authors (T.M.A) is grateful to Prof. Dr. S. Randjbar--Daemi for
the hospitality extended to him at the Abdus Salam International Center for
theoretical Physics, Trieste, Italy, where parts of this work are carried  
out.

\newpage

\section*{Figure captions}
{\bf Fig. (1)} The dependence of the asymmetry parameter $\alpha$ on $q^2$ 
for the $\Lambda_b \rar \Lambda \mu^+ \mu^-$ decay, at five different fixed
values of the vector type Wilson coefficient $C_{RR}$.\\ \\
{\bf Fig. (2)} The same as in Fig. (1), but for the coefficient 
$C_{RL}$. \\ \\
{\bf Fig. (3)} The dependence of the asymmetry parameter $\alpha_\theta$ on 
$q^2$ for the $\Lambda_b \rar \Lambda \mu^+ \mu^-$ decay, at five different 
fixed values of the vector type Wilson coefficient $C_{LL}$. \\ \\
{\bf Fig. (4)} The same as in Fig. (3), but for the coefficient
$C_{LR}$. \\ \\
{\bf Fig. (5)} The dependence of the asymmetry parameter $\beta_\theta$ on 
$q^2$ for the $\Lambda_b \rar \Lambda \tau^+ \tau^-$ decay, at five different 
fixed values of the vector type Wilson coefficient $C_{LR}$.\\ \\
{\bf Fig. (6)} The dependence of the asymmetry parameter $\alpha_{\Lambda_S}$ 
on $q^2$ for the $\Lambda_b \rar \Lambda \mu^+ \mu^-$ decay, at five 
different fixed values of the vector type Wilson coefficient $C_{LL}$. \\ \\
{\bf Fig. (7)} The same as in Fig. (6), but for the coefficient 
$C_{RL}$. \\ \\
{\bf Fig. (8)} The same as in Fig. (6), but for $\Lambda_b \rar \Lambda
\tau^+ \tau^-$ decay. \\ \\
{\bf Fig. (9)} The same as in Fig. (8), but for the coefficient 
$C_{LR}$.\\ \\
{\bf Fig. (10)} The same as in Fig. (8), but for the coefficient 
$C_{RL}$.\\ \\
{\bf Fig. (11)} The same as in Fig. (8), but for the coefficient 
$C_{RR}$.

\newpage

\begin{figure}
\vskip 1.5 cm
    \includegraphics{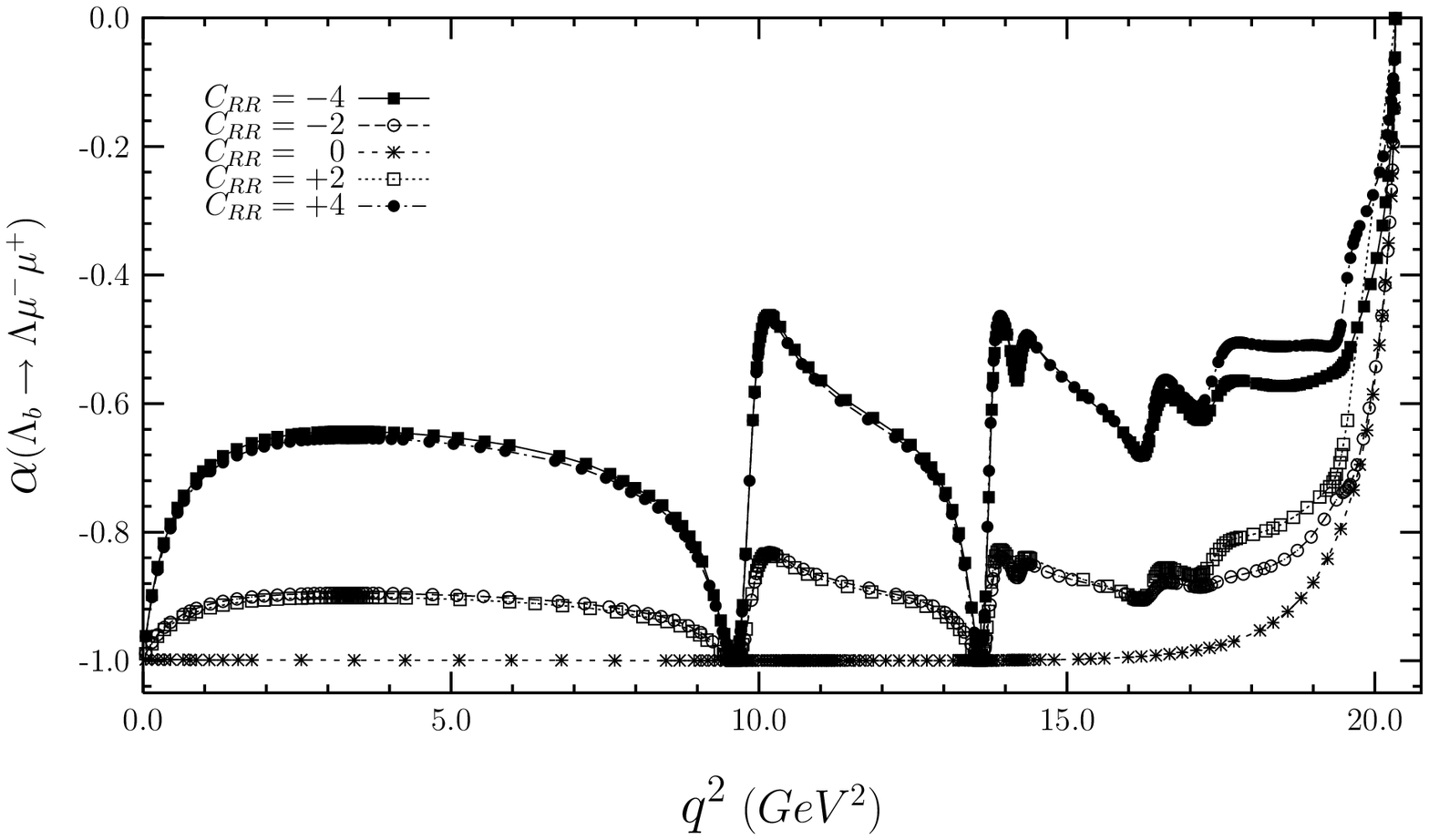}
\vskip 7.8cm
\caption{}
\end{figure}

\begin{figure}
\vskip 2.5 cm
    \includegraphics{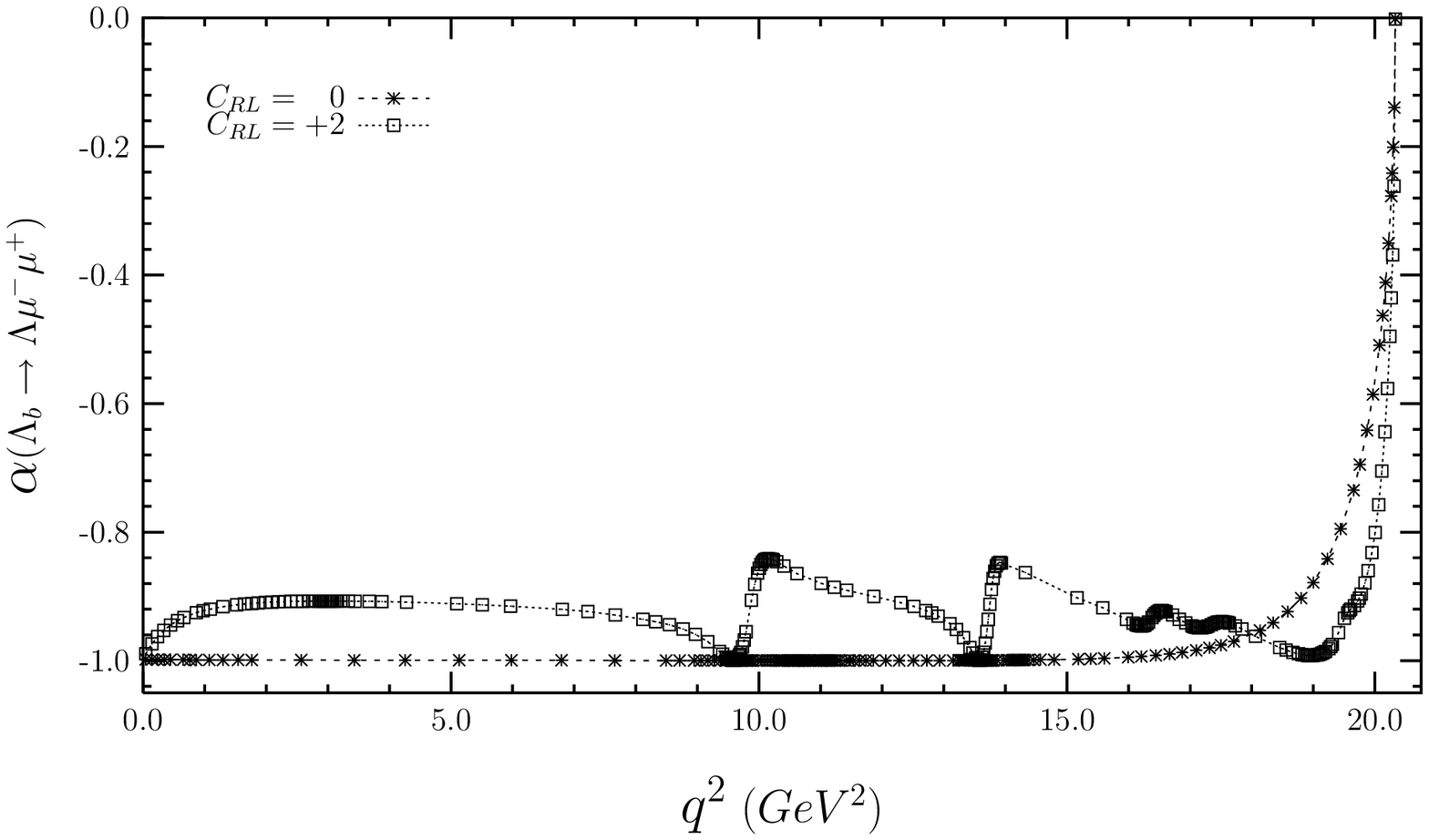}
\vskip 7.8 cm
\caption{}
\end{figure}

\begin{figure}
\vskip 1.5 cm
    \includegraphics{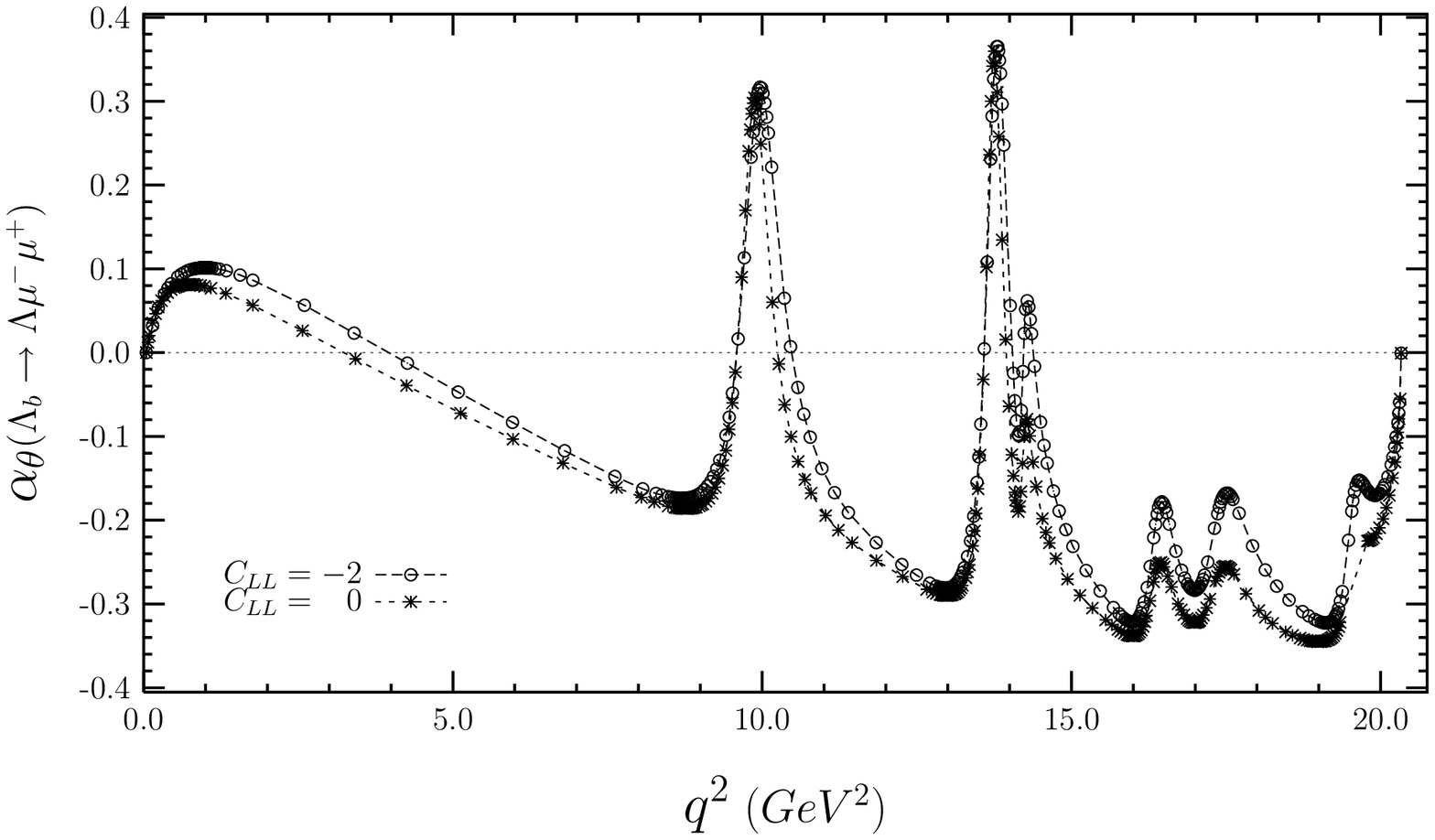}
\vskip 7.8cm
\caption{}
\end{figure}

\begin{figure}
\vskip 2.5 cm
    \includegraphics{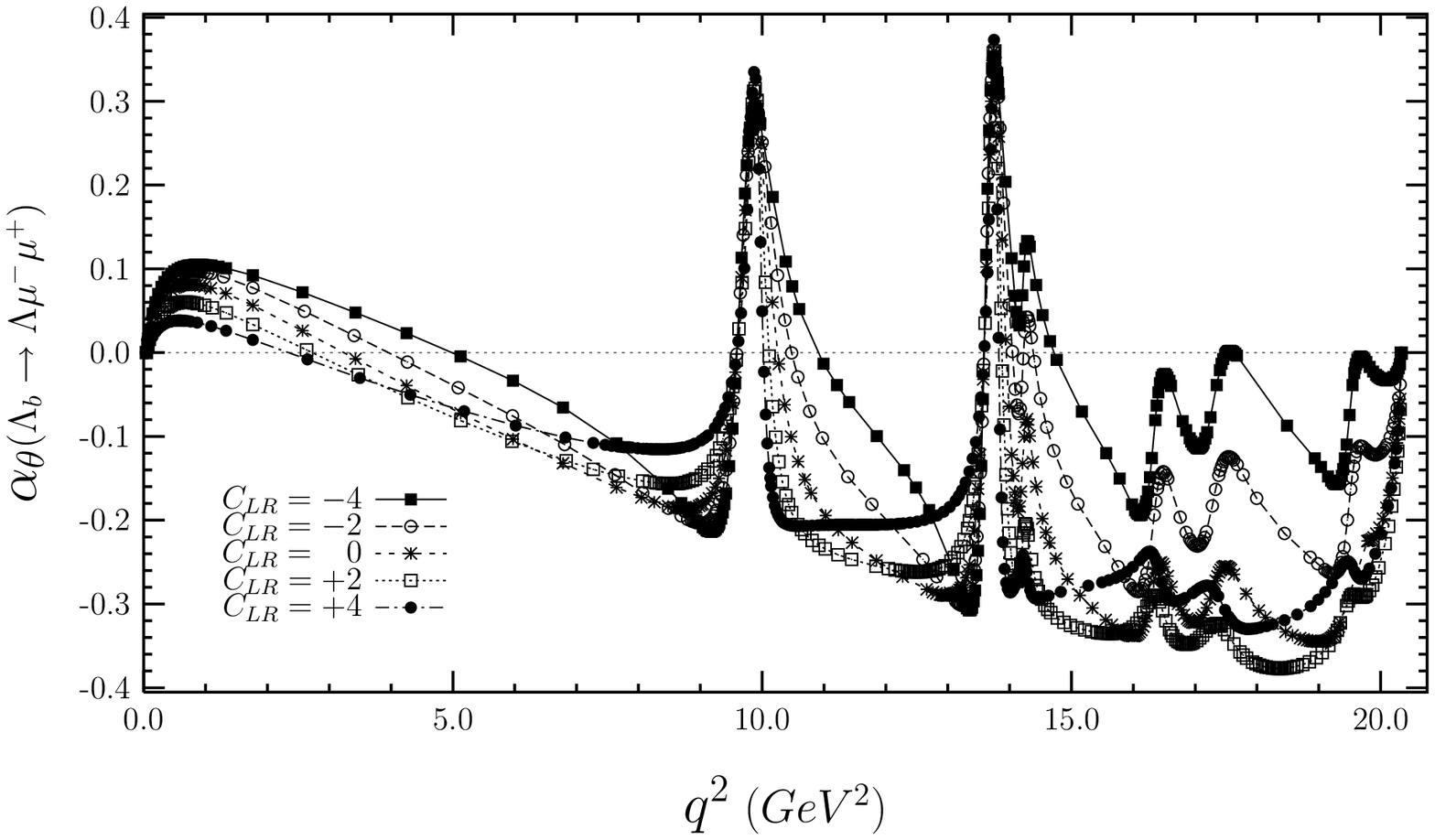}
\vskip 7.8 cm
\caption{}
\end{figure}

\begin{figure}
\vskip 2.5 cm
    \includegraphics{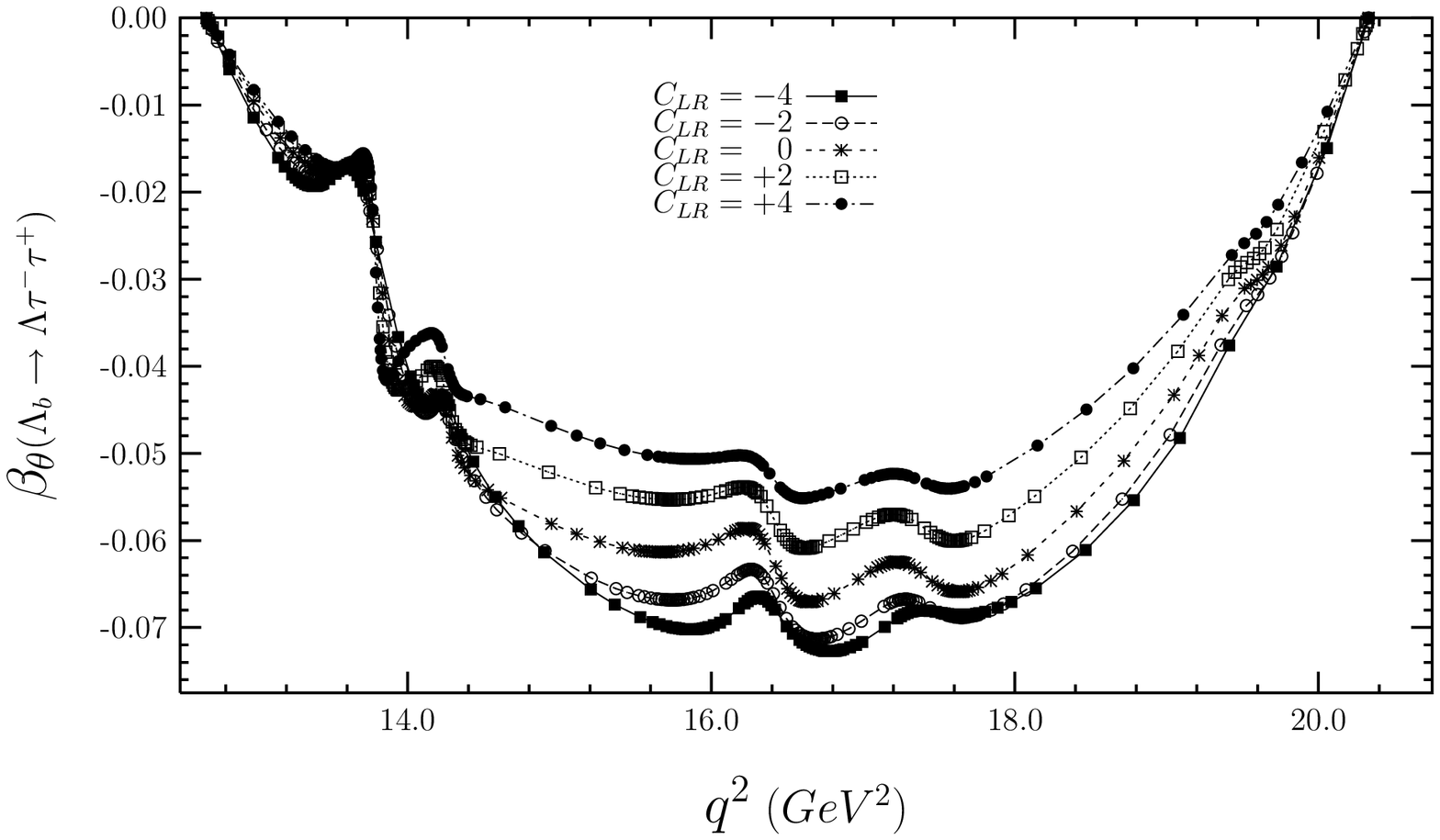}
\vskip 7.8 cm
\caption{}
\end{figure}

\begin{figure}
\vskip 1.5 cm
    \includegraphics{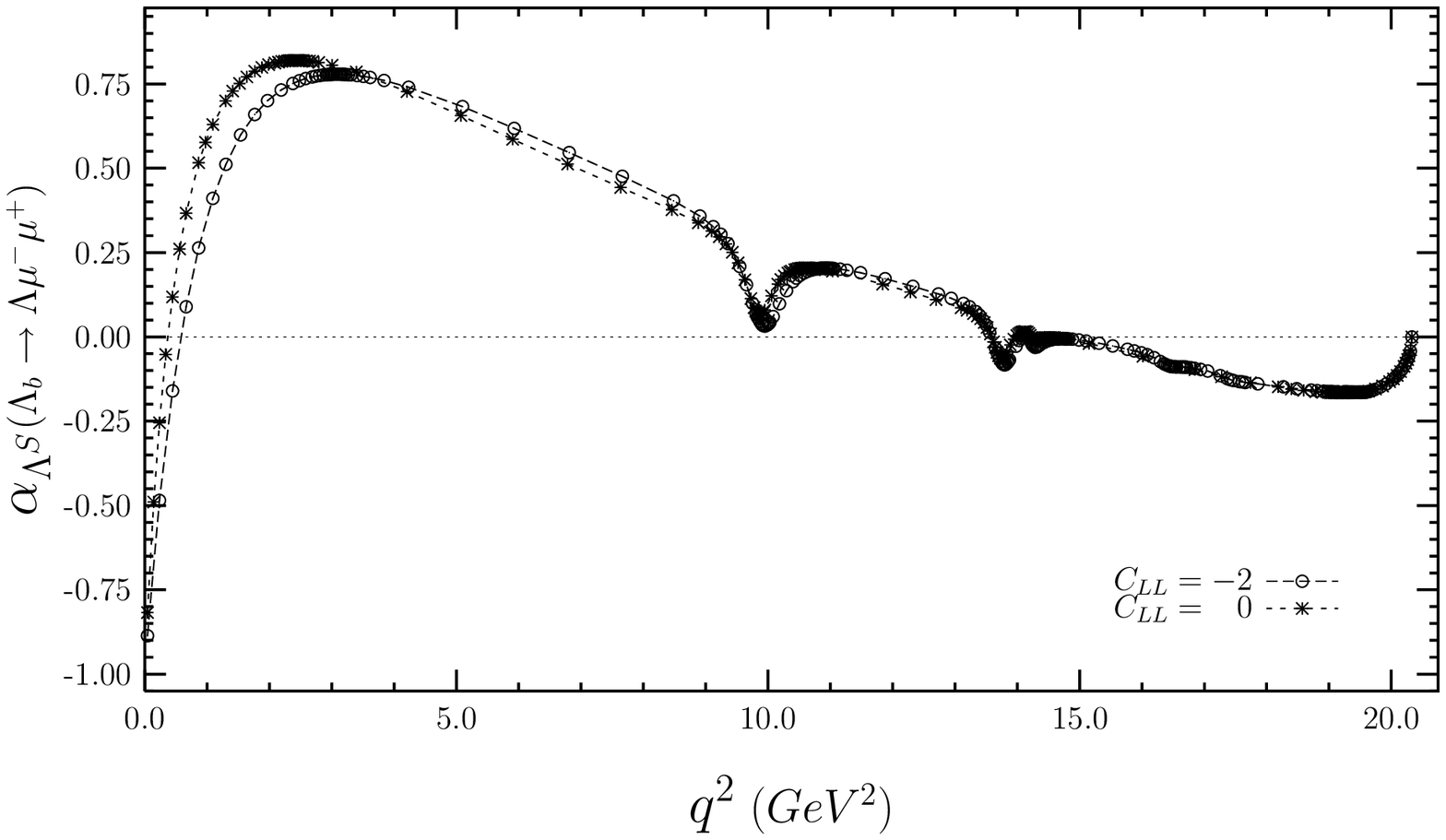}
\vskip 7.8cm
\caption{}
\end{figure}

\begin{figure}
\vskip 2.5 cm
    \includegraphics{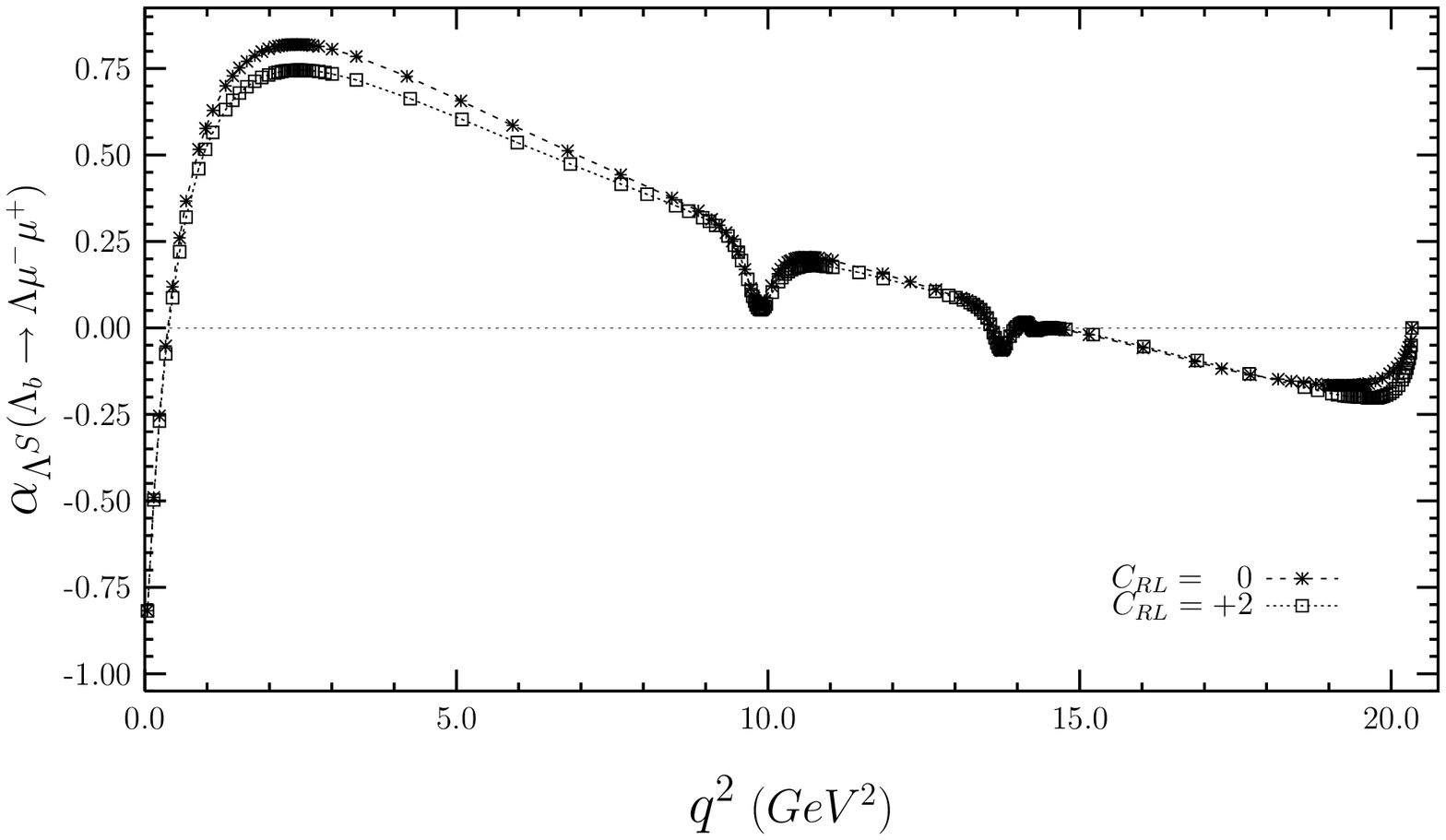}
\vskip 7.8 cm
\caption{}
\end{figure}

\begin{figure}
\vskip 1.5 cm
    \includegraphics{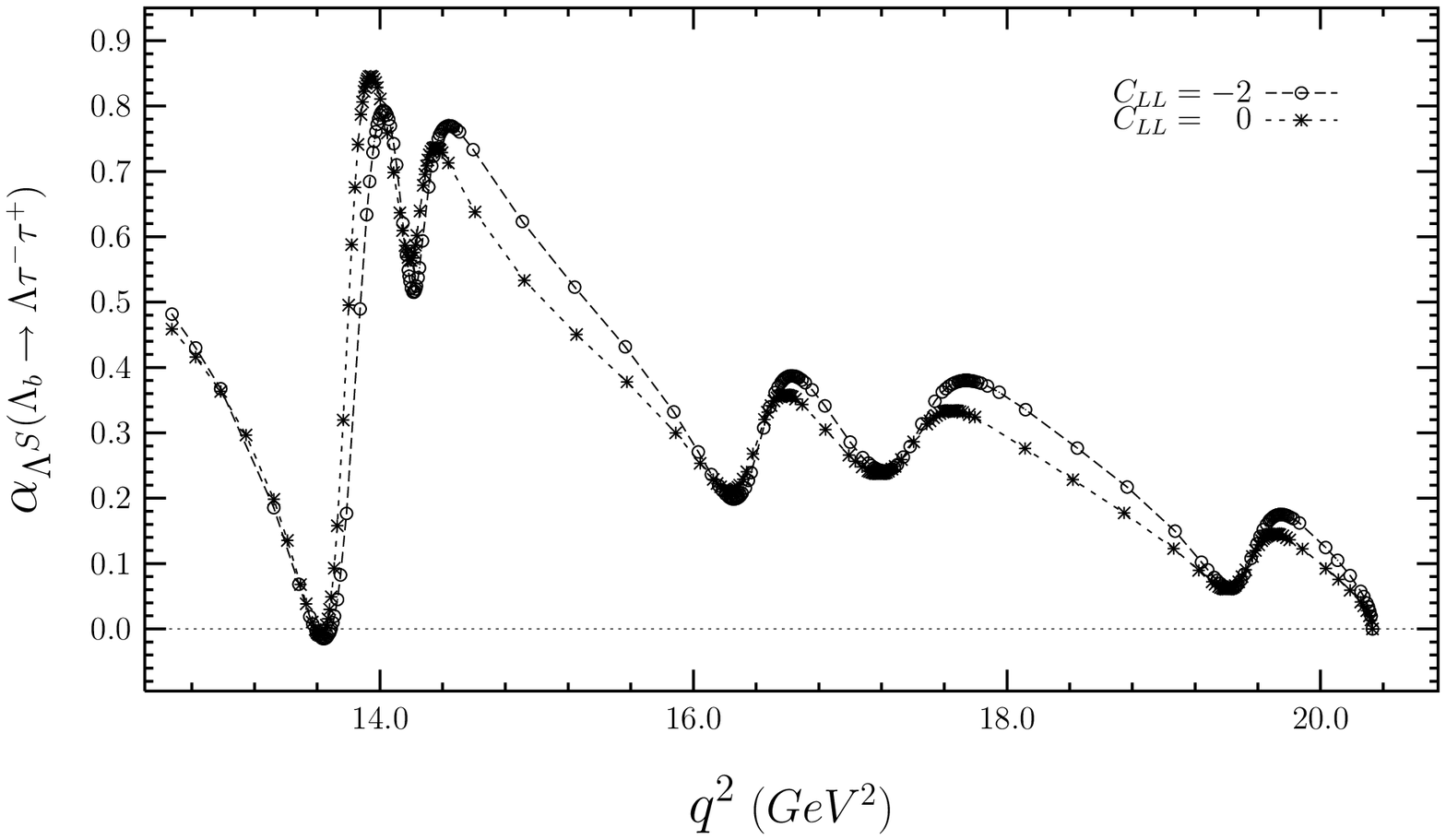}
\vskip 7.8cm
\caption{}
\end{figure}

\begin{figure}
\vskip 2.5 cm
    \includegraphics{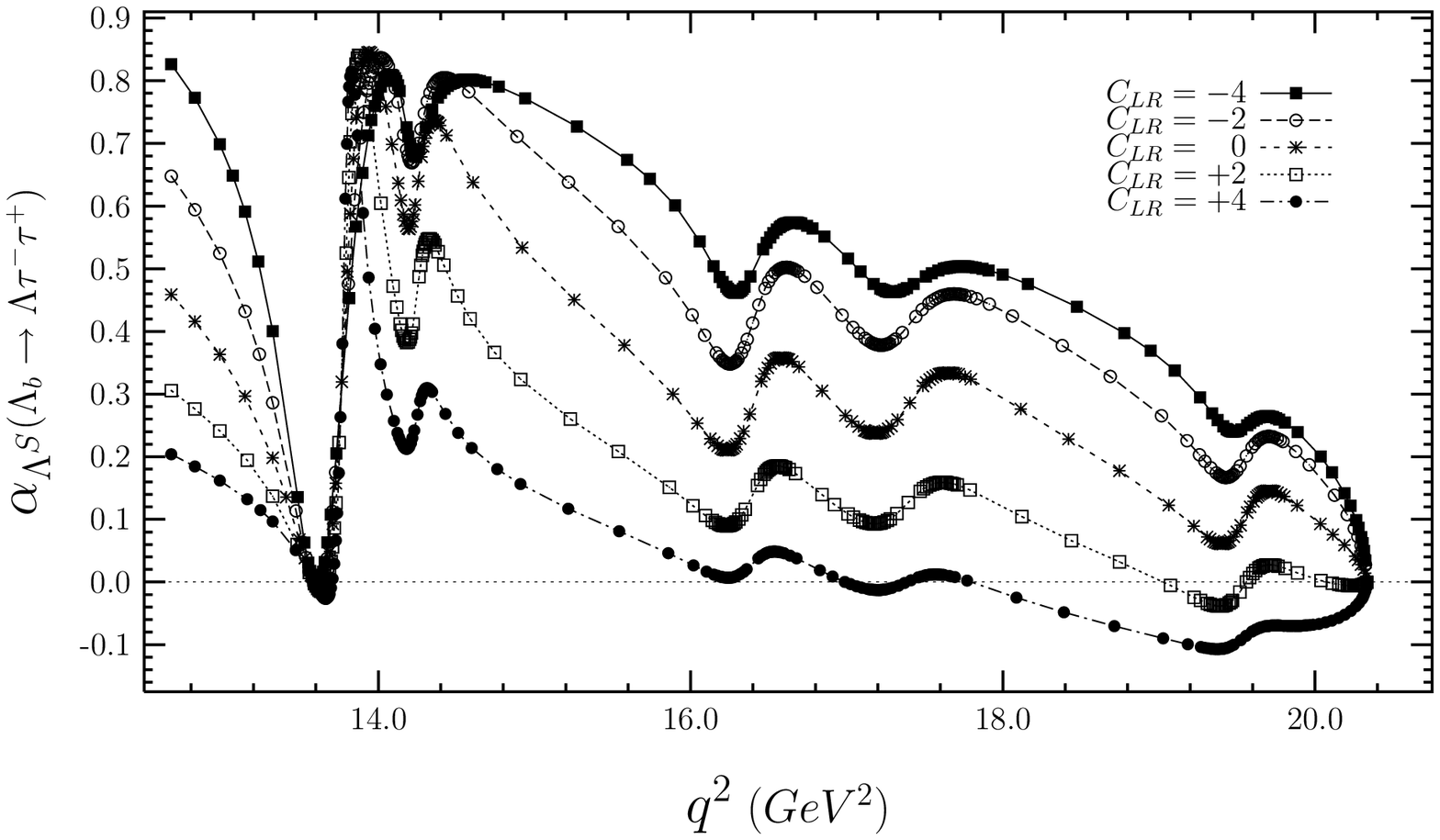}
\vskip 7.8 cm
\caption{}
\end{figure}

\begin{figure}
\vskip 1.5 cm
    \includegraphics{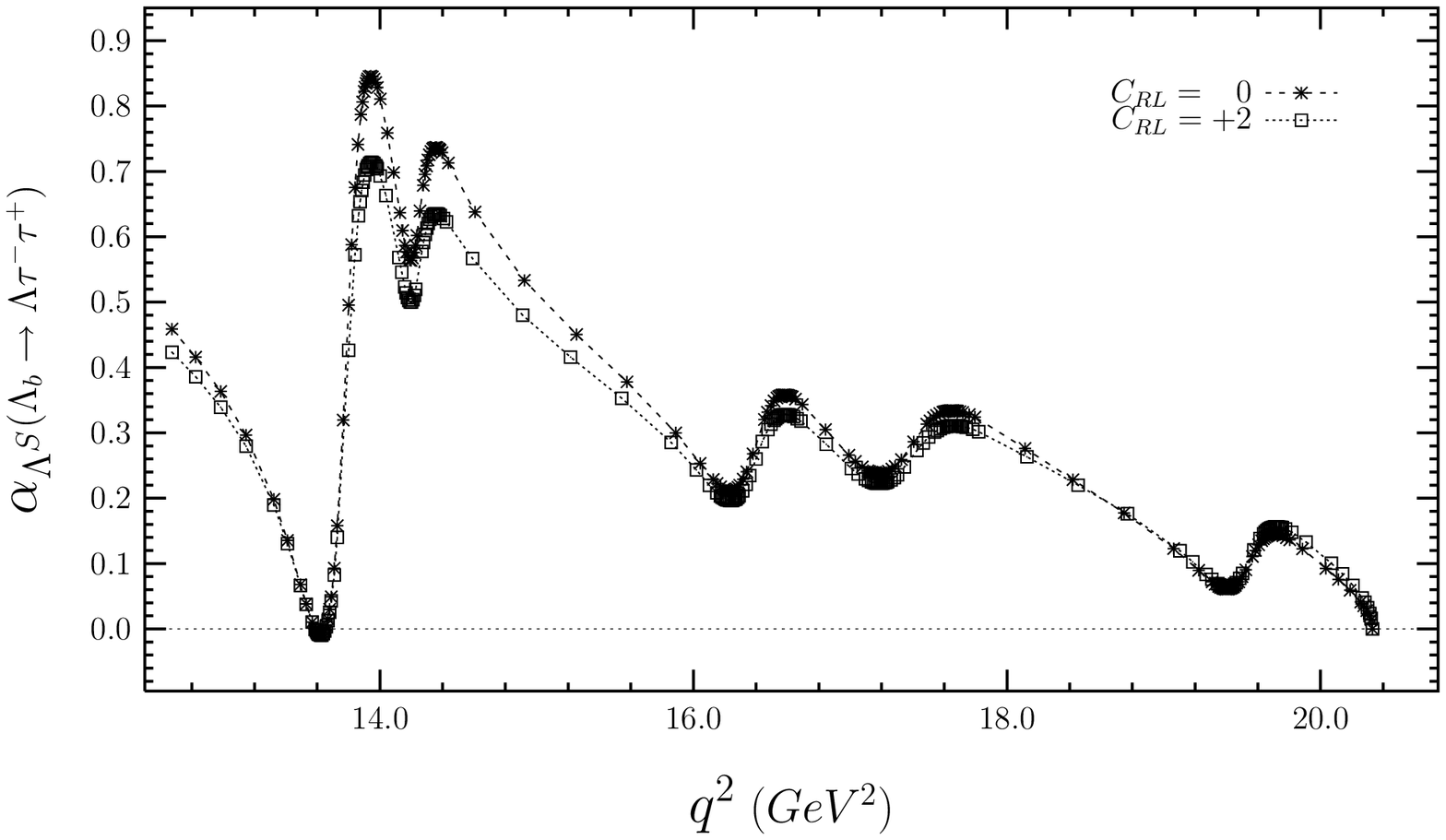}
\vskip 7.8cm
\caption{}
\end{figure}

\begin{figure}
\vskip 2.5 cm
    \includegraphics{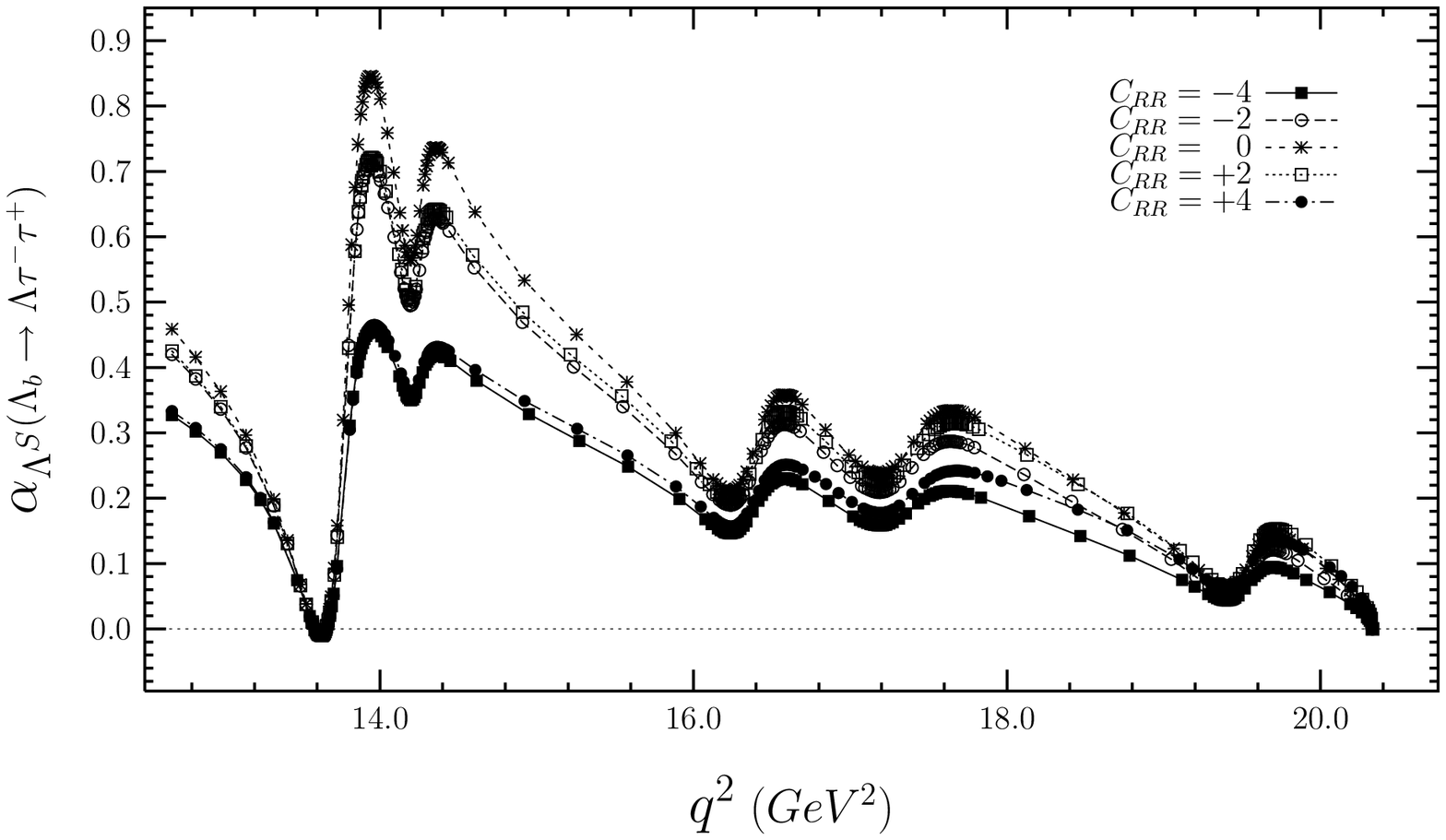}
\vskip 7.8 cm
\caption{}
\end{figure}

\end{document}